\documentclass{iopart}

\usepackage{latexsym}

\usepackage{setstack}

\usepackage[titlenumbered,ruled,noend,english,onelanguage]{algorithm2e}

\SetCommentSty{mycommfont}
\SetEndCharOfAlgoLine{}

\usepackage{harvard}
\usepackage{iopams}
\usepackage{scrextend}

\usepackage[english]{babel}
\usepackage{amsthm}
\usepackage{dsfont}
\usepackage{color}
\usepackage{float}

\usepackage{graphicx}
\usepackage{subfig}

\graphicspath{{images/},{prebuiltimages/},{revision/}}

\usepackage{LpHypermodels}

\usepackage{cleveref} 
\crefname{pluralequation}{eqs.}{eqs.}
\Crefname{pluralequation}{Eqs.}{Eqs.}
\crefformat{equation}{#2eq.~(#1)#3}
\Crefformat{equation}{#2Eq.~(#1)#3}

\newlength{\tempwidth}
\newcommand{\columnname}[1]
{\makebox[\tempwidth][c]{\textbf{#1}}}

\begin{document}

\title[Majorization-minimization and hierarchical Bayesian modeling for M/EEG]{A hierarchical Bayesian perspective on majorization-minimization for non-convex sparse regression: application to M/EEG source imaging}


\author{Yousra Bekhti$^{1}$, Felix Lucka$^{2,3}$, Joseph Salmon$^{1}$, Alexandre Gramfort$^{1,4}$}

\address{
$^1$ LTCI, T\'el\'ecom ParisTech, Universit\'e Paris-Saclay, Paris, France.\\
$^2$ Centrum Wiskunde \& Informatica, Science Park 123, 1098 XG Amsterdam, Netherlands\\
$^3$ Centre for Medical Image Computing, University College London, WC1E 6BT London, UK\\
$^4$ INRIA, Parietal team, Universit\'e Paris-Saclay, Saclay, France.
}
\ead{yousra.bekhti@telecom-paristech.fr}

\begin{abstract}

Majorization-minimization (MM) is a standard iterative optimization technique which consists in minimizing a sequence of convex surrogate functionals. MM approaches have been particularly successful to tackle inverse problems and statistical machine learning problems where the regularization term is a sparsity-promoting concave function. However, due to non-convexity, the solution found by MM depends on its initialization.
Uniform initialization is the most natural and often employed strategy as it boils down to penalizing all coefficients equally in the first MM iteration.
Yet, this arbitrary choice can lead to unsatisfactory results in severely under-determined inverse problems such as source imaging with magneto- and electro-encephalography (M/EEG).
The framework of hierarchical Bayesian modeling (HBM) is an alternative approach to encode sparsity.
This work shows that for certain hierarchical models, a simple alternating scheme to compute fully Bayesian maximum \emph{a posteriori} (MAP) estimates leads to the exact same sequence of updates as a standard MM strategy (cf. the Adaptive Lasso).
With this parallel outlined, we show how to improve upon these MM techniques
by probing the multimodal posterior density using Markov Chain Monte-Carlo (MCMC) techniques.
Firstly, we show that these samples can provide well-informed initializations that help MM schemes to reach better local minima.
Secondly, we demonstrate how it can reveal the different modes of the posterior distribution in order
to explore and quantify the inherent uncertainty and ambiguity of such ill-posed inference procedure.
In the context of M/EEG, each mode corresponds to a plausible configuration of neural sources, which is crucial for data interpretation, especially in clinical contexts. Results on both simulations and real datasets show how the number or the type of sensors affect the uncertainties on the estimates.

\end{abstract}

\maketitle


\section{Introduction}
\label{sec:introduction}

Over the last two decades, sparsity has emerged as a key concept to solve inverse problems such as tomographic image reconstruction, deconvolution or inpainting, but also to regularize high dimensional regression problems in the field of machine learning. There are mainly two routes to introduce sparsity in such problems.\\

The first route, embraced by the optimization community and frequentist
statisticians, is to promote sparsity using convex optimization theory.
This line of work has led to now mature theoretical guarantees~\cite{FoRa13} when using regularization functions based on $\ell_1$-norm and other convex variants~\cite{Tibshirani96}. In particular, it has been popularized in the signal processing community under the name of compressed sensing~\cite{candes2008introduction} when combined with incoherent measurements.

There are however some limitations of sparsity-promoting convex penalties based
on the $\ell_1$-norm.
All the features (also called regressors, atoms or sources depending
on the terminology of the community) involved
in the solution form what is called the support of the solution.
Convex penalties can fail to identify the correct support in the presence of highly noisy data, but also in low noise setups if the forward operator (referred to as design matrix in statistics) is poorly conditioned.
Convex regularizations also lead to a systematic underestimation bias
in the amplitude of the coefficients~\cite{OsBuGoXuYi06,Candes,chartrand2007exact,saab2008stable,ChHeSa17}.

To address these limitations of $\ell_1$-type models,
re-weighted schemes have been
proposed~\cite{Candes,Gasso,Rakotomamonjy,zhang-rao:2011,strohmeier-etal:16}, of which the Adaptive Lasso~\cite{Zou06} is the most commonly used in the statistics community:
starting from the Lasso estimator, which amounts to regressing with a standard $\ell_1$-norm as a regularizer (this estimator is sometimes referred to as Basis Pursuit Denoising (BPDN)~\cite{Chen_Donoho_Saunders98} in signal processing), the Adaptive Lasso solves a sequence of weighted Lasso problems, where at each iteration the
weights are chosen such that the strongest coefficients are less and less penalized.
From the optimization point of view, such an iterative scheme can be derived from so-called majorization-minimization (MM) strategies~\cite{lange2000optimization,schifano2010majorization}.
The idea behind MM is to minimize the objective function by successively minimizing upper bounds that are easier to optimize. Many well-known optimization approaches can be interpreted as instances of MM, \eg simple gradient descent or proximal algorithms~\cite{Combettes2011}, expectation-maximization (EM)~\cite{Dempster77maximumlikelihood}, and difference-of-convex (DC) programming techniques~\cite{Horst:1999}.
More recently, re-weighted $\ell_1$-norm schemes based on MM principle have been particularly popular to handle concave, hence non-convex regularizations such as $\ell_{0.5}$-quasi-norms or logarithmic functions. As such, these schemes are prone to converging to a local minimum determined by the initial, uniformly weighted $\ell_1$-norm solution (\ie the Lasso estimator) that constitutes the first iterate.\\

The common way to formulate HBMs is to consider the variance parameters of Gaussian prior models as additional random variables which have to be estimated from the data as well. Their prior distributions are referred to as hyper-priors. Plausible solutions to the regression problem, that both fit data and the \emph{a priori} assumption of sparsity, are explicitly characterized as multiple distinct modes of the posterior distribution. This characterization is the Bayesian analogue to local minima in variational regression approaches when working with non-convex functionals. To infer a point estimate of the parameters of interest from the \emph{a posteriori} distribution different strategies exist. For instance Variational Bayesian approaches~\cite{mackay2003information,jordan1999introduction,sato2004hierarchical,FrHaDaKiPhTrHeFlMa08,shervashidze2015learning}, Sparse Bayesian Learning (SBL) approaches (also referred to as type-I or type-II maximum likelihood estimates) \cite{tipping2001sparse,wipf2004sparse,Wipf-Nagarajan:2009,zhang-rao:2011} and fully-Bayesian strategies \cite{CaHaPuSo09,Lucka-etal:2012} are possible.
In this work, we focus on the latter one for a non-standard type of HBM examined in~\cite{Lu14} that combines a non-Gaussian prior with an $\ell_{1}$-type energy function with a specific Gamma hyper-prior.
Interestingly, for this HBM, a simple alternating scheme to compute full maximum \emph{a posteriori} (MAP) estimates leads to exactly the same sequence of problems solved by MM applied to $\ell_{1/2}$-type regularizations. In other words, the Adaptive Lasso estimator~\cite{Zou06} commonly used in machine learning is tightly related to this HBM model.
With this observation made, it is natural to revisit and improve these MM schemes by leveraging the ability of the Bayesian framework to explore the modes of the posterior distribution by Markov chain Monte-Carlo (MCMC) schemes~\cite{RoCa05,KaSo05}. This can not only mitigate the aforementioned initialization-dependence of MM, but more importantly, it can offer insights into the structure and importance of potentially multiple plausible sparse solutions. Yet, the benefit comes at the cost of additional computational efforts.

Magnetoencephalography and electroencephalography (M/EEG) are technologies that allow to measure the electromagnetic fields produced by active neurons in a non-invasive way. Localization of foci of neural activations from M/EEG recordings is a high impact problem both for cognitive neuroscience and clinical neuroscience, with applications in pathologies such as epilepsy, sleep research or neurodegenerative disorders. Despite the linearity of the forward problem, this inverse problem is
particularly challenging as the forward operator is both under-determined and strongly
ill-conditioned.
As such, both non-convex optimization strategies with re-weighted schemes~\cite{strohmeier-etal:16} and hierarchical Bayesian approaches have been proposed~\cite{sato2004hierarchical,CaHaPuSo09,Wipf-Nagarajan:2009,Sorrentino-etal:2009,Lucka-etal:2012} for M/EEG source localization. For this reason, it is an ideal application for our examinations.\\

The manuscript is organized as follows: first, we present a unified
perspective on both routes to sparsity, \ie re-weighted $\ell_1$ MM schemes
and specific HBMs. We show that a particular optimization-based inference strategy recovers the MM algorithm. We then describe a HBM inference strategy based upon an MCMC sampling and show on simulated and experimental M/EEG datasets how these stochastic MCMC-based techniques can not only help to improve upon deterministic approaches, but also help to reveal multiple plausible solutions to the inverse problem. This analysis leads to an uncertainty quantification (UQ) of the support recovery of non-convex sparse regression problems that provides very useful complementary information, in particular for very ill-conditioned and under-determined applications like M/EEG source localization.


\subsection*{Notation}
\label{sec:notation}

We consider here the following linear model, known in machine learning as multi-task regression and in signal processing as the multiple measurement vector (MMV) model~\cite{cotter-etal:2005}:
\begin{equation} \label{eq:FwdEq}
    \bfM = \bfG \bfX + \bfE \enspace,
\end{equation}
where $\bfM\in\R^{m\times t}$.
In machine learning $m$ corresponds to the number of samples
and $t$ to the number of tasks, and
in our application $m$ corresponds to the number of sensors
and $t$ to the number of measurements over time.
The matrix $\bfG\in\R^{m\times q}$ is the design matrix, a known instantaneous mixing matrix also referred to as the \emph{forward}, \emph{gain} or \emph{system} matrix. It relates the unknown coefficients $\bfX\in\R^{q \times t}$, which correspond to amplitudes of neural sources, to the measurements $\bfM$. In our application $q=dn$ where $n$ is the number of source locations and $d$ is the number of oriented sources per location ($d=1$ or $d=3$ corresponding to estimating a scalar field or a 3D vector field of sources).
The matrix $\bfE$ models the measurement noise, which is assumed to be an \termabb{additive, white Gaussian noise}{AWGN}. This is a reasonable assumption after performing a proper spatial whitening of the data using an estimate of the noise covariance~\cite{engemann2015automated}.
By $\bfX_{(i,j)}$ we refer to the entry in the $i$-th row and $j$-th column in $\bfX$, while $\bfX_{(i,:)} \in \R^{t}$ and $\bfX_{(:,j)} \in \R^{q}$ refer to the complete $i$-th row and $j$-th column, respectively. In addition, we denote by $\bfX_{[i]}$, $i = 1,\ldots,n$ the $d \times t$ sub-matrix of $\bfX$ corresponding to the $i$-th \emph{group}: for $d=1$, this coincides with $\bfX_{(i,:)}$, while for $d=3$, $\bfX_{[i]} = [\bfX_{((i-1)d+1,:)}^\top,\bfX_{((i-1)d+2,:)}^\top,\bfX_{((i-1)d+3,:)}^\top]^\top$. 
Note that thereby, the group size is $dt$. Negative indices are used to exclude certain entries, rows, columns or groups, \ie, $\bfX_{(:,-j)} \in\R^{q \times (t-1)}$ refers to the matrix obtained by deleting the $j$-th column of $\bfX$. Furthermore, let $I_m$ denote the identity matrix of size $m \in \bbN$. For any matrix $\bfA\in \bbR^{n \times m}$, the Frobenius norm is given by $\fronormsq{\bfA}=\sum_{i,j} \bfA_{(i,j)}^2$.


\section{Methods}
\label{sec:methods}

We start this section by recalling how majorization-minimization works
when addressing variational formulations with concave, non-convex, regularization.
It is followed by an introduction to hierarchical Bayesian models with Gamma hyper-priors.
Then, we explain how these seemingly different approaches can lead to the
exact same optimization algorithm.
From this, we detail how different Bayesian inference strategies using MCMC
sampling can more precisely explore the landscape of the posterior distribution of the HBM model,
as well as provide multiple possible solutions to the sparse regression problem.

\subsection{Majorization-Minimization}
\label{sub:MM}

Majorization-Minimization (MM) strategies consist in replacing a difficult
optimization problem with a series of easier ones that are obtained by
upper bounding the objective function, often by a convex majorant.
In the context of inverse problems or high-dimensional statistics using sparsity constraints,
MM has been successfully applied to address non-convex regularization terms.
An example is the regression model with $\ell_{2,p}$-quasi-norms regularization when $0<p<1$. The desired estimate $\hat{\bfX}$ is defined as one of potentially multiple minimizers of
\begin{equation} \label{eq:L2pReg}
\hat{\bfX} \in \argmin_{\bfX\in\R^{q \times t}} \frac{1}{2} \fronormsq{\bfM - \bfG \bfX}  + \lambda \sum_{i=1}^n \fronorm{\bfX_{[i]}}^p \enspace,
\end{equation}
where $\lambda > 0$ is the regularization parameter balancing the data fit and the penalty term. One possible MM approach to solve Eq.~\eref{eq:L2pReg} with $p=1/2$ would consist in minimizing a sequence of non-smooth convex surrogate functions where the non-convex regularization is replaced by a weighted $\ell_{2,1}$ norm~\cite{strohmeier-etal:16}. In each iteration, the weights are derived from the current estimate of $\bfX$.

Due to the concavity of the non-decreasing function $\bfX\mapsto\sqrt{\fronorm{\bfX}}$, it is upper bounded by its tangent and a first order Taylor expansion at the current estimate $\bfX_{[i]}$ provides an upper bound that can be used to construct the non-smooth convex surrogate problem. By solving this sequence of surrogate problems, the value of the non-convex objective function is guaranteed to decrease. However, due to the non-convexity, only convergence towards a local minimum can be guaranteed.

For the problem in Eq.~\eref{eq:L2pReg} with $p=1/2$, the $k$-th iteration of the MM scheme reads:
\begin{eqnarray}
\label{eq:MM}
\fl \quad \hat{\bfX}^{(k)} \in \argmin_{\bfX\in\R^{q \times t}} \frac{1}{2} \fronormsq{\bfM - \bfG \bfX}  + \lambda \sum_{i=1}^n \frac{ \fronorm{\bfX_{[i]}} }{ w^{(k-1)}_{i}}, \qquad w^{(k-1)}_{i} = 2 \sqrt{\fronorm{\bfX^{(k-1)}_{[i]}}} \enspace.
\end{eqnarray}
As each weight $w^{(k)}_{i}$ is a non-decreasing function of $\fronorm{\bfX^{(k)}_{[i]}}$, sources with high amplitudes in one iteration will be less penalized in the next iteration. Strong sources are more and more promoted to explain the data $\bfM$. Sources for which $\fronorm{\bfX^{(k)}_{[i]}} = 0$ at a certain iteration $k$ are effectively pruned from the model for all following iterations. Using MM therefore leads to a solution that explains the data with fewer active  locations $i$ compared to a standard $\ell_{2,1}$ norm regularized solution.
Note that a default initialization consists in setting $w^{(0)}_{i}=1, \forall i \in [n]$.

To exploit existing fast solvers for the $\ell_{2,1}$ regularized problems~\cite{strohmeier-etal:16,Ndiaye_Fercoq_Gramfort_Salmon15}, we reformulate the weighted subproblem and apply the weights by scaling the matrix $\bfG$ with a diagonal matrix $\bfW^{(k)} \in \bbR^{dn \times dn}$ given by:
\begin{eqnarray} \label{eq:weights}
\bfW^{(k)} = \mathrm{diag}(\bfw^{(k)} \otimes \mathbf{1}_{d}) \enspace ,
\end{eqnarray}
where $\bfw^{(k)} \in \bbR^{n}$, $\mathbf{1}_{d}\in\R^d$ is a vector of ones and $\otimes$ is the Kronecker product.
Defining $\tilde{\bfG}^{(k)}=\bfG \bfW^{(k-1)}$, the reformulated problem reads:
\begin{eqnarray}\label{eq:lasso_new_gram}
\tilde{\bfX}^{(k)} &
 & \in \argmin_{\bfX\in\R^{q \times t}} \frac{1}{2}\fronormsq{\bfM - \tilde{\bfG}^{(k)} \bfX}  + \lambda \sum_{i=1}^n \fronorm{\bfX_{[i]}} \enspace .
\end{eqnarray}
After convergence, we reapply the scaling to $\tilde{\bfX}$ to obtain $\hat{\bfX}$:
\begin{equation}
    \label{eq:MM_weights}
    \hat{\bfX}^{(k)} = \bfW^{(k-1)} \tilde{\bfX}^{(k)} \enspace .
\end{equation}
The reformulation through Eq.~\eref{eq:lasso_new_gram} and \eref{eq:MM_weights} avoids any division by zero when $\bfX^{(k-1)}=0$. The above procedure, which matches the strategy of the Adaptive Lasso estimator~\cite{Zou06}, is expressed as pseudo-code in Algorithm~\ref{alg:adpative_lasso}. More technical details can be found in~\cite[Algorithm 3]{strohmeier-etal:16}.

{\fontsize{4}{4}\selectfont
\begin{algorithm}[t]
\SetKwInOut{Input}{input}
\SetKwInOut{Init}{init}
\SetKwInOut{Parameter}{param}
\caption{\textsc{$\ell_{2,p}$ MM algorithm with $p=1/2$ (Adaptive Lasso)}}
\Input{$\bfM, \bfG,\lambda > 0, \bfW^{(0)} \geqslant 0,\epsilon > 0, \tau > 0$ and $K$ }
\For{
        $k = 1$ to $K$
    }
    {
		$\tilde{\bfG}^{(k)} = \bfG \bfW^{(k-1)}$

		Get $\tilde{\bfX}^{(k)}$ solving Eq.~\eref{eq:lasso_new_gram} at $\epsilon$-precision (e.g. by block coordinate descent).

		Update $ \hat{\bfX}^{(k)} = \bfW^{(k-1)} \tilde{\bfX}^{(k)}$

	    Update $\bfW^{(k)}=\mathrm{diag}(\bfw^{(k)} \otimes \mathbf{1}_{d})$ where $\bfw^{(k)}_{[i]} = 2 \sqrt{\fronorm{\hat{\bfX}^{(k)}_{[i]}}}$,  $\forall i\in [n]$

		\If{ $\norm{\hat{\bfX}^{(k)}-\hat{\bfX}^{(k-1)}}_{\infty} \leq \tau$}{Break}

     }
\Return{$\hat{\bfX}^{(k)}$}
\label{alg:adpative_lasso}
\end{algorithm}
}
\subsection{Hierarchical Bayesian Modeling}
\label{sub:HBM}

In this section, we formulate the inference problem given by Eq.~\eref{eq:FwdEq} and the regularization strategy with $\ell_{2,p}$-quasi-norms from a Bayesian perspective \cite{KaSo05,Lu14}: the Bayesian approach incorporates prior beliefs about the model parameters in terms of probability distributions. Under the AWGN assumption the likelihood of the model is given by:
\begin{eqnarray} \label{eq:like}
\like &\propto \exp \left( - \frac{1}{2} \fronormsq{\bfM - \bfG \bfX} \right) \enspace.
\end{eqnarray}
From Eq.~\eref{eq:L2pReg} we can construct the $\ell_{2,p}$ group prior as:
 \begin{equation} \label{eq:prior}
\prior \propto \exp \left( - \lambda \sum_{i=1}^n \fronorm{\bfX_{[i]}}^p \right)
= \prod_{i=1}^n \exp \left( - \lambda \fronorm{\bfX_{[i]}}^p \right) \enspace,
\end{equation}
which leads to the following posterior probability density using Bayes rule:
\begin{equation} \label{eq:post}
\post \propto \exp \left( - \frac{1}{2} \fronormsq{\bfM - \bfG \bfX} - \lambda \sum_{i=1}^n \fronorm{\bfX_{[i]}}^p \right) \enspace.
\end{equation}
To extend Eq.~\eref{eq:prior} to a hierarchical prior model \cite{mackay2003information}, we replace the scalar $\lambda$ by a vector of hyper-parameters $\gamma \in \R^{n}_+$ and for any $p \geq 1$ we write the \emph{conditional $\ell_{2,p}$ prior} as:
\begin{equation} \label{eq:condprior}
\fl \hiprior
\propto \exp \left( - \sum_{i=1}^n \left( \frac{\fronorm{\bfX_{[i]}}^p}{\gamma_i} + \frac{d t}{p} \log(\gamma_i)\right)\right) \enspace,
\end{equation}
%
where the logarithmic term accounts for the terms of the normalization that depend on $\gamma$ \cite{Lu14}. A common choice for a hyper-prior on each $\gamma_i$ is given by a \emph{Gamma distribution} \cite{mackay2003information,KaSo05,CaHaPuSo09,Lucka-etal:2012} with shape and scale parameters $\alpha$ and $\beta$:
\begin{equation} \label{eq:hyper}
\fl \hyper \propto
\prod_{i=1}^n \gamma_i^{\alpha - 1}
\exp \left(- \frac{\gamma_i}{\beta} \right)
=\exp \left(\sum_{i=1}^n \left( - \frac{\gamma_i}{\beta} + (\alpha - 1) \log(\gamma_i) \right) \right) \enspace.
\end{equation}
Then, the full posterior over both $\bfX$ and $\gamma$ becomes:
\begin{eqnarray}
\label{eq:full-post}
\fl \hipost \propto \nonumber \\
\fl \hspace{3em} \exp \left( - \frac{1}{2} \fronormsq{\bfM - \bfG \bfX} - \sum_{i=1}^n \left( \frac{\fronorm{\bfX_{[i]}}^p}{\gamma_i} + \frac{\gamma_i}{\beta} - (\alpha - 1 - \frac{dt}{p}) \log(\gamma_i) \right) \right)\enspace.
\end{eqnarray}
The question of how to best derive parameter estimates, in particular how to treat the two different types of parameters $\bfX$ and $\gamma$, distinguishes different HBM-based inference strategies. Variational Bayesian approaches ~\cite{mackay2003information,jordan1999introduction,sato2004hierarchical,FrHaDaKiPhTrHeFlMa08,shervashidze2015learning} and Sparse Bayesian Learning \cite{tipping2001sparse,wipf2004sparse,Wipf-Nagarajan:2009,zhang-rao:2011} approaches rely on approximating or marginalizing the full, joint posterior distribution \eref{eq:full-post}. In contrast, fully-Bayesian strategies \cite{CaHaPuSo09,Lucka-etal:2012} work with it directly. The most popular one is the full maximum-a-posteriori (\emph{full-MAP}) estimate which is defined as
\begin{eqnarray}
(\xMAP,\gamMAP) & \in \argmax_{(\bfX,\gamma) \in\R^{q \times t} \times \R^{n}_+} \left\lbrace \hipost \right\rbrace \enspace.
\end{eqnarray}
A common strategy  to compute it is to minimize the \emph{negative log posterior} $-\log \hipost$ by alternating  minimization over $\bfX$ and $\gamma$ (known as \emph{block coordinate descent} in optimization):
\begin{eqnarray}
\label{eq:AO}
\fl \qquad  \bfX^{(k)}& \in \argmin_{\bfX\in\R^{q \times t}} \left\lbrace \frac{1}{2} \fronormsq{\bfM - \bfG \bfX} + \sum_{i=1}^n  \frac{\fronorm{\bfX_{[i]}}^p}{\gamma^{(k-1)}_i} \right\rbrace \label{eq:AO-X}\enspace, \\
\fl \qquad \gamma^{(k)}_i & \in \argmin_{\gamma_i\in \R_+} \left\lbrace\frac{\fronorm{\bfX^{(k)}_{[i]}}^p }{\gamma_i} + \frac{\gamma_i}{\beta} - (\alpha - 1 - \frac{dt}{p}) \log(\gamma_i) \right\rbrace, \quad \forall i\in [n] \enspace. \label{eq:AO-gamma}
\end{eqnarray}
Other fully-Bayesian estimates are defined as integrals of functions of $\bfX$ and $\gamma$ with respect to the posterior distribution, \eg first or second moment estimates. To compute these high dimensional integrals efficiently, only Markov chain Monte Carlo (MCMC) methods that draw correlated samples from the posterior distribution can be used~\cite{RoCa05,KaSo05}. A commonly used MCMC scheme for HBM is given by \emph{blocked Gibbs sampling} which alternates as:
\begin{eqnarray}
\bfX^{(k)} &\sim \: p_{post}(\bfX,\gamma^{(k-1)}|\bfM) &\propto p_{post}(\bfX|\bfM,\gamma^{(k-1)}) \enspace, \label{eq:AS-X}\\
\gamma^{(k)} &\sim \: p_{post}(\bfX^{(k)},\gamma|\bfM) &\propto p_{post}(\gamma|\bfM,\bfX^{(k)})\enspace. \label{eq:AS-gamma}
\end{eqnarray}
In this study, however, we are not interested in sampling the posterior distribution to compute the integral-based estimators but we rather want to explore the different modes of this multi-modal distribution, each of which corresponds to parameters that are both sparse and likely to explain the data.\\
One can notice similar structures in \eref{eq:AO-X}-\eref{eq:AO-gamma} and \eref{eq:AS-X}-\eref{eq:AS-gamma}. In each step, we make use of the conditional structure of the posterior: for $\gamma$ fixed, we have to solve one $qt$-dimensional $\ell_{2,p}$ optimization/sampling problem, while for $\bfX$ fixed, we have to solve $n$ 1-dimensional optimization/sampling problems. We will describe these two steps in more detail in the next two sections.

\subsection{HBM Optimization}
\label{sub:optimization}

The optimization problem defined in Eq.~\eref{eq:AO-X} reduces to an $\ell_{2,p}$-norm regularized regression problem that can be solved as described in Section \ref{sub:MM}. To solve Eq.~\eref{eq:AO-gamma}, we compute the first order optimality condition for each $i$:
\begin{eqnarray}
\label{eq:OptiGamma}
\qquad - \frac{\fronorm{\bfX^{(k)}_{[i]}}^p}{\gamma_i^2} + \frac{1}{\beta} - \frac{( \alpha - 1 - \frac{dt}{p})}{\gamma_i} &= 0 \enspace,\label{eq:OptiGammaFirst}
\end{eqnarray}
For $\alpha \geqslant d t/p + 1$, the problem in Eq.~\eref{eq:AO-gamma} is convex, and the positive root of Eq.~\eref{eq:OptiGammaFirst} is given by:
\begin{equation}
\gamma_i = \beta \left( \nu + \sqrt{ \nu^2 + \frac{\fronorm{\bfX^{(k)}_{[i]}}^p}{\beta}} \right), \qquad \nu \mydef \frac{\alpha - 1 - dt/p}{2} \enspace.
\end{equation}
Note that similar rules to update the noise level were considered in the Bayesian Lasso \cite{Park_Casella08,Kyung_Gill_Ghosh_Casella10} and the Scaled Lasso (see for instance \cite{Stadler_Buhlmann_vandeGeer10,Dalalyan12}). A difference though is that the update we perform here is on the penalty term, whereas in the mentioned references, it was rather performed on the data-fitting term.

If we furthermore choose $\alpha = d t/p + 1$, then $\nu = 0$ and most terms disappear; Eq.~\eref{eq:AO-X} and \eref{eq:AO-gamma} hence read:
\begin{eqnarray}
\bfX^{(k)} &= \argmin_{\bfX\in\R^{q \times t}} \left\lbrace \frac{1}{2} \fronormsq{\bfM -\bfG \bfX} + \sum_{i=1}^n  \frac{\fronorm{\bfX_{[i]}}^p}{\gamma^{(k-1)}_i} \right\rbrace \enspace, \label{eq:AO-nu0-X}\\
\gamma^{(k)}_i &= \sqrt{\beta} \sqrt{\fronorm{\bfX^{(k)}_{[i]}}^p} \, , \quad \forall i=1,\ldots,n \enspace, \label{eq:AO-nu0-gamma}
\end{eqnarray}
which can be combined to the fixed point iteration:
\begin{equation} \label{eq:AO-nu0-2}
\bfX^{(k)} = \argmin_{\bfX\in\R^{q \times t}} \left\lbrace \frac{1}{2} \fronormsq{\bfM -\bfG \bfX} + \frac{2}{\sqrt{\beta}} \sum_{i=1}^n  \frac{\fronorm{\bfX_{[i]}}^p}{2 \sqrt{\fronorm{\bfX^{(k-1)}_{[i]}}^p}} \right\rbrace \enspace .
\end{equation}
If we compare Eq.~\eref{eq:AO-nu0-2} with Eq.~\eref{eq:MM} we see that we re-derived the MM algorithm for $p=1$ as an alternating optimization scheme to compute the \emph{full-MAP} estimate for a specific HBM, namely using a conditional $\ell_{2,1}$ group prior and a Gamma hyper-prior with $\alpha = dt + 1$ and $\beta = 4/\lambda^2$. Using $\bfw^{(0)}_{i} := 1$ in the MM scheme corresponds to starting with $\gamma_i^{(0)} := 1/\lambda =  \sqrt{\beta}/2$.
From previous work~\cite{strohmeier-etal:16} we know that due to the non-convexity, a good initialization of the weights $\bfw^{(0)}_{i}$ in the MM algorithm is crucial for its performance, but aside from uniform initialization, only heuristic initialization strategies were used, \eg using the same re-weighting as in the sLORETA method \cite{Pa02}. In this work, we leverage the re-interpretation of the MM algorithm through the HBM framework to obtain multiple initializations in a systematic fashion, namely as samples drawn from the full posterior. This way, we can not only reach better local minima but more importantly, we can identify and characterize multiple possible sparse solutions. Such plausible solutions to the sparse regression problem in Eq.~\eref{eq:FwdEq} are the modes of the posterior distribution \eref{eq:full-post} with different relative probability masses.

\subsection{HBM Sampling}
\label{sub:sampling}

As outlined in Eq.~\eref{eq:AS-X} and \eref{eq:AS-gamma} in Section~\ref{sub:HBM}, we sample the full posterior $\hipost$ by blocked Gibbs sampling, \ie we alternate between sampling the conditional distributions $p_{post}(\bfX|\bfM,\gamma^{(k-1)})$ and $p_{post}(\gamma|\bfM,\bfX^{(k)})$. The conditional $p_{post}(\bfX|\bfM,\gamma^{(k-1)})$ is a high dimensional distribution composed of a Gaussian likelihood and an $\ell_{2,p}$ prior, where our main interest here is $p = 1$. It was demonstrated in \cite{Lu12} that \termabb{single component Gibbs sampling}{SC Gibbs} is an efficient MCMC technique to sample such distributions. For the specific $\ell_{2,p}$ priors used here, \emph{slice sampling} can be used to perform the sub-steps in SC Gibbs sampling, namely the sampling of the one-dimensional single-component conditional densities. The resulting \emph{Slice-Within-Gibbs} sampler was examined in \cite{Lu16}. For completeness, the details of the implementation are given in \ref{sec:DetailsSliceSampler}.\\
The conditional $p_{post}(\gamma|\bfM,\bfX^{(k)})$ factorizes over groups $i$:
\begin{equation}
\fl \qquad p_{post}(\gamma_i|\bfM,\bfX^{(k)}) \propto \exp \left( -\frac{\fronorm{\bfX^{(k)}_{[i]}}^p}{\gamma_i} - \frac{\gamma_i}{\beta} + (\alpha - 1 - d t/p) \log(\gamma_i) \right) \enspace. \label{eq:conPostGammai}
\end{equation}
For the case of $\alpha = d t/p + 1$, which is our main interest due to its connection to MM revealed in the previous section, Eq. \eref{eq:conPostGammai} reduces to:
\begin{equation}
\fl \qquad p_{post}(\gamma_i |\bfM, \bfX^{(k)}) \propto \exp \left(- \frac{\fronorm{\bfX^{(k)}_{[i]}}^p}{\gamma_i} \right) \exp \left(- \frac{\gamma_i}{\beta} \right) \enspace, \label{eq:conPostGammaiRed}
\end{equation}
which can be sampled with a simple accept-reject algorithm as described in \ref{sec:DetailsGammaSampler}. The complete procedure is described in Algorithm~\ref{alg:sampling}. Therein, $K_0$ refers to the burn-in size, \ie the initial samples that are discarded, $K$ to the sample size of the blocked Gibbs sampler. We denote $K_{SC}$ and $K_{SS}$ the sample sizes of the SC Gibbs and the slice sampler that carry out the sampling in the sub-steps.

{\fontsize{4}{4}\selectfont
\begin{algorithm}[t]
\SetKwInOut{Input}{input}
\SetKwInOut{Init}{init}
\SetKwInOut{Parameter}{param}
\caption{\textsc{Block Gibbs Sampling scheme}}
\Input{$\bfM, \bfG $, $\bfX^{(-K_0)}$, $\gamma^{(-K_0)}$, $K_0$, K, $K_{Gibbs}$, $K_{SC}$, $K_{SS}$, $\alpha$, $\beta$}

\For{
        $k = -K_0+1$ to $K$
    }
    {
    Set $\bfX^{(k)} = \bfX^{(k-1)}$.\\
    \For{
    		$k_{SC} = 1$ to $K_{SC}$
    		}
    		{
    		Draw a random permutation $P$ of $\{1,\ldots,n\}$\\
    		\For{
    			$l \in P$
    			}
    			{
			Sample $\bfX_{(i,j)}^{(k)} \sim p_{post}(\bfX_{(i,j)}|\bfX_{-(i,j)}^{(k-1)},\bfM,\gamma^{(k)})$, $\forall (i,j)\in [l]$ - via $K_{SS}$ steps of \textbf{Slice Sampling} Algorithm \ref{alg:Slice}.
			}
		}
	{Sample $\gamma^{(k)}_{i} \sim p_{post}(\gamma_{i}|\bfM,\bfX^{(k)})$, $\forall i =1,\ldots,n$ via \textbf{Accept-Reject} Algorithm \ref{alg:acceptreject}.}

    }
\Return{$\{\bfX^{(k)},\gamma^{(k)}\}_{k=1}^K$}
\label{alg:sampling}
\end{algorithm}
}

\subsection{Combining Sampling and Optimization}
\label{sub:sampling_opti}

Finding the correct support in a sparse under-determined regression problem like \eref{eq:FwdEq} is inherently of combinatorial complexity.
In the two approaches we examined, this is reflected in the non-convexity of the objective function \eref{eq:L2pReg} and the multi-modality of the joint posterior distribution \eref{eq:full-post}, respectively.
Here, we want to investigate whether the link between MM and the HBM framework can be used to quantify the ambiguity and uncertainty posed by sparse support identification.
Traditional uncertainty quantification (UQ) measures such as covariance estimates of $\bfX$ or $\gamma$ may fail to do so as they cannot capture the multi-modality of the posterior distribution in a satisfactory way.
In addition, no sample $\bfX^{(k)}$ is exactly sparse: as the posterior distribution is a continuous density function, the event $\{\bfX^{(k)}_{[i]} = 0\}$ has zero probability.
This means that the whole support of $\bfX^{(k)}$ is active with probability $1$.
Even a thresholded average of the support of $\bfX^{(k)}$ will only reveal the average probability of a location being part of the support. In our application to M/EEG source analysis, an arguably more interesting statistical output is the set sources forming the network of brain areas active in a given data set. This question needs a more profound spatial analysis of the structure of the most prominent modes of the posterior, and is left open by the above mentioned measures. Here, we propose to tackle it in a different way: Algorithm~\ref{alg:sampling_optim} describes a combination of first sampling the posterior and then using each sample to initialize MM to optimize the posterior distribution.
This yields a chain of different posterior modes $\{\hat{\bfX}^{(k)}\}_{k=1}^K$, \ie approximate solutions to Eq.~\eref{eq:FwdEq} that fulfill our \emph{a priori} knowledge of a sparse support. If we assume that the division of $\R^{q\times t}$ into attractors of the MM algorithm roughly overlaps with the division of $\R^{q\times t}$ into modes of the marginalized posterior over $\bfX$ within the HBM framework, the relative frequency with which these modes occur in $\{\hat{\bfX}^{(k)}\}_{k=1}^K$ corresponds to their relative posterior mass. While a mathematically more profound and detailed analysis of this heuristic is left for future work, we will illustrate in the following numerical examples how this mode analysis can be used to reveal and quantify some of the ambiguity of sparse under-determined regression problems.

{\fontsize{4}{4}\selectfont
\begin{algorithm}[ht]
\SetKwInOut{Input}{input}
\SetKwInOut{Init}{init}
\SetKwInOut{Parameter}{param}
\caption{\textsc{Combination of Gibbs sampler and MM algorithm}}
\Input{$\bfM, \bfG $, $\lambda$, $\bfX^{(-K_0)}$, $\gamma^{(-K_0)}$, $K_0$, $K$, $K_{SC}$, $K_{SS}$, $K_{MM}$, $\epsilon > 0$, $\tau > 0$}

Use Algorithm~\ref{alg:sampling} with input\\ ($\bfM$, $\bfG $, $\bfX^{(-K_0)}$, $\gamma^{(-K_0)}$, $K_0$, $K$, $K_{SC}$, $K_{SS}$, $\alpha = dt +1 $, $\beta = 4/\lambda^2$) to obtain MCMC chain $\{\bfX^{(k)},\gamma^{(k)}\}_{k=1}^K$.\\
\For{
        $k = 1$ to $K$
    }
    {
	 Set $\bfw^{(0)}_{i} = \lambda \gamma_i^{(k)}, \forall i=1,\dots,n$ and run Algorithm~\ref{alg:adpative_lasso} with input
	 ($\bfM, \bfG,\lambda, \bfW^{(0)}_i, \epsilon, \tau$, $K_{MM}$) to obtain $\hat{\bfX}^{(k)}$.
    }
\Return{$\{\hat{\bfX}^{(k)}, \bfX^{(k)},\gamma^{(k)}\}_{k=1}^K$}
\label{alg:sampling_optim}
\end{algorithm}
}

\section{Results}
\label{sec:Res}
We now examine the benefits of our re-interpretation of the MM algorithm described in Section \ref{sub:MM} as a specific way to compute a full-MAP estimate for a specific HBM as described in Sections \ref{sub:HBM} and \ref{sub:optimization}.
We first illustrate basic properties of the methods in a one dimensional toy problem before we examine a simulated MEG dataset and two experimental M/EEG datasets.

\subsection{One Dimensional Illustrations}

\begin{figure}[htp]
	\centering
	\includegraphics[width = 1\textwidth]{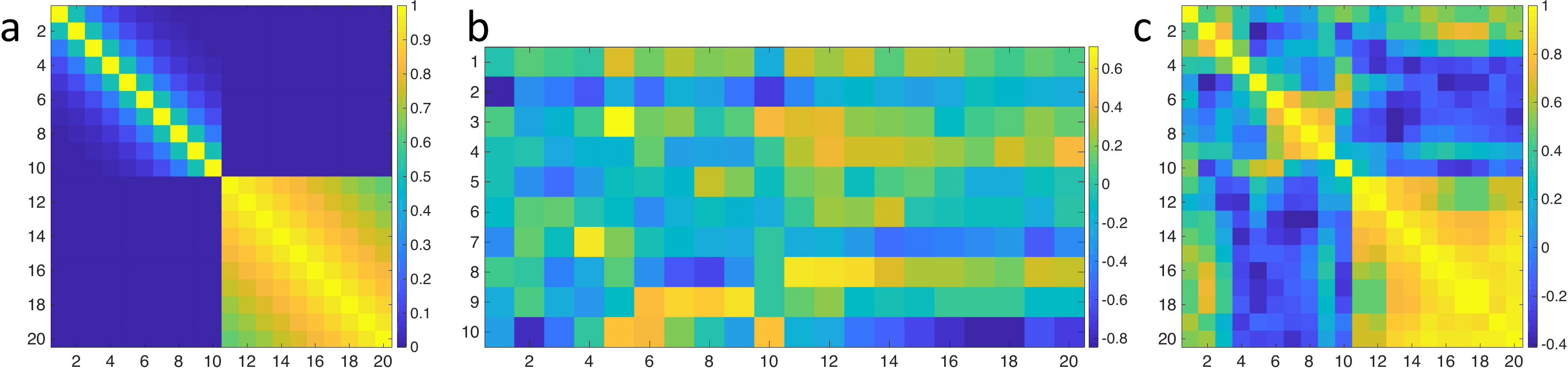}
	\caption{Setup for the first one dimensional illustration.  a) Random Gaussian design covariance matrix $\mathbf{C}$; b) design matrix $\bfG$; c) $\bfG^T \bfG$.}
	\label{fig:1DExa1Setup}
\end{figure}

\begin{figure}[htp]
	\centering
	\includegraphics[width = 0.7\textwidth]{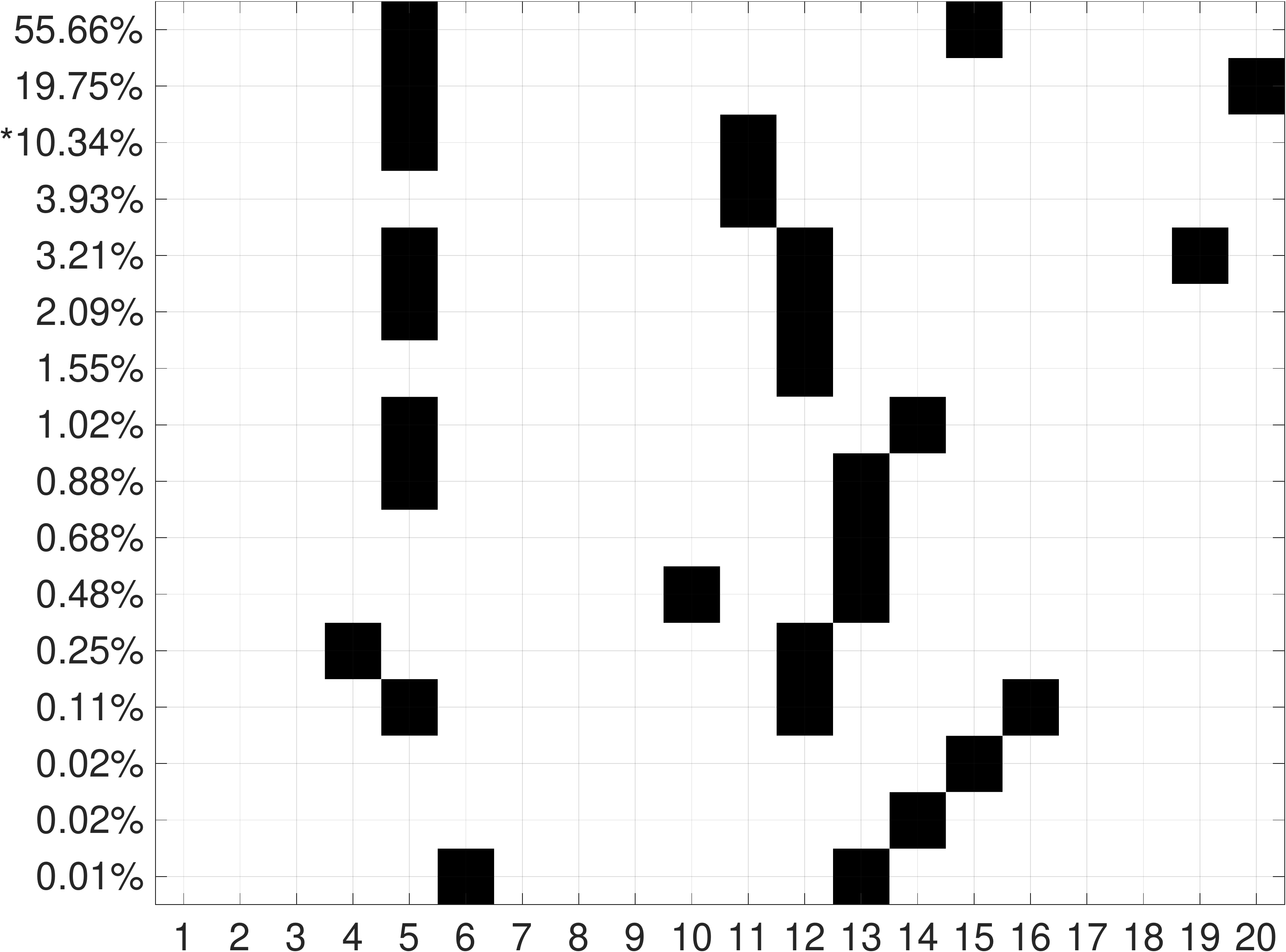}

	\caption{Each row of the matrix indicates the spatial support of a mode found by Algorithm~\ref{alg:sampling_optim} in Example 1 (the true locations are $5$ and $15$). The rows are ordered by the relative frequency (in $\%$), by which they are found. The row marked by * corresponds to the mode found by MM with uniform initialization.}
	\label{fig:1DExa1ModeFreq}
\end{figure}

\begin{figure}[htp]
	\centering
	\includegraphics[width = \textwidth]{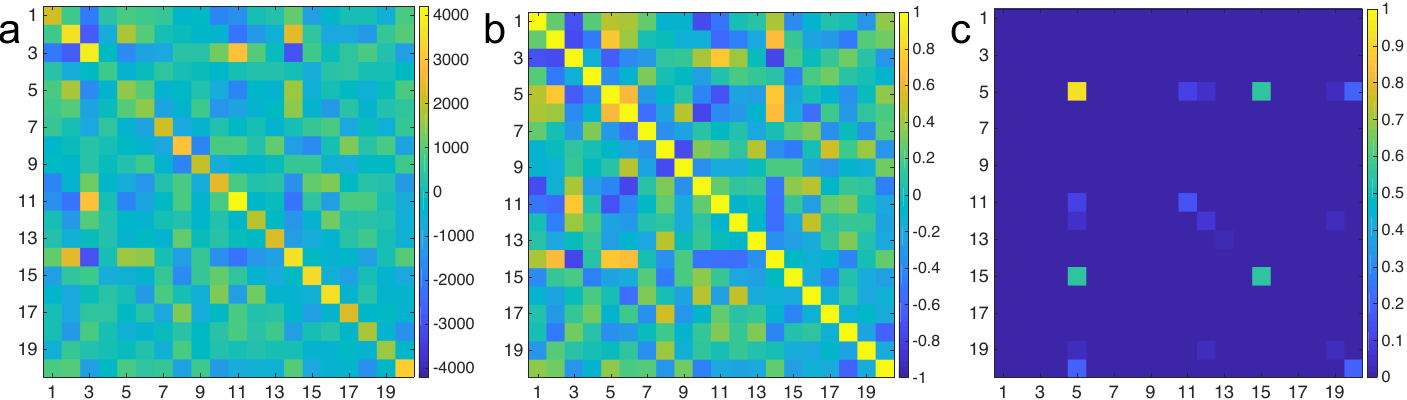}

	\caption{Comparison between a) posterior sample covariance matrix ; b) corresponding correlation matrix; (c) a matrix which shows the frequency with with locations $i$ and $j$ are simultaneously found active in the $\hat{\bfX}^{(k)}$ in its $(i,j)^{th}$ entry.}
	\label{fig:1DExa1CovVsModeMaps}
\end{figure}

\begin{figure}[htp]
	\centering
	\includegraphics[width = 1\textwidth]{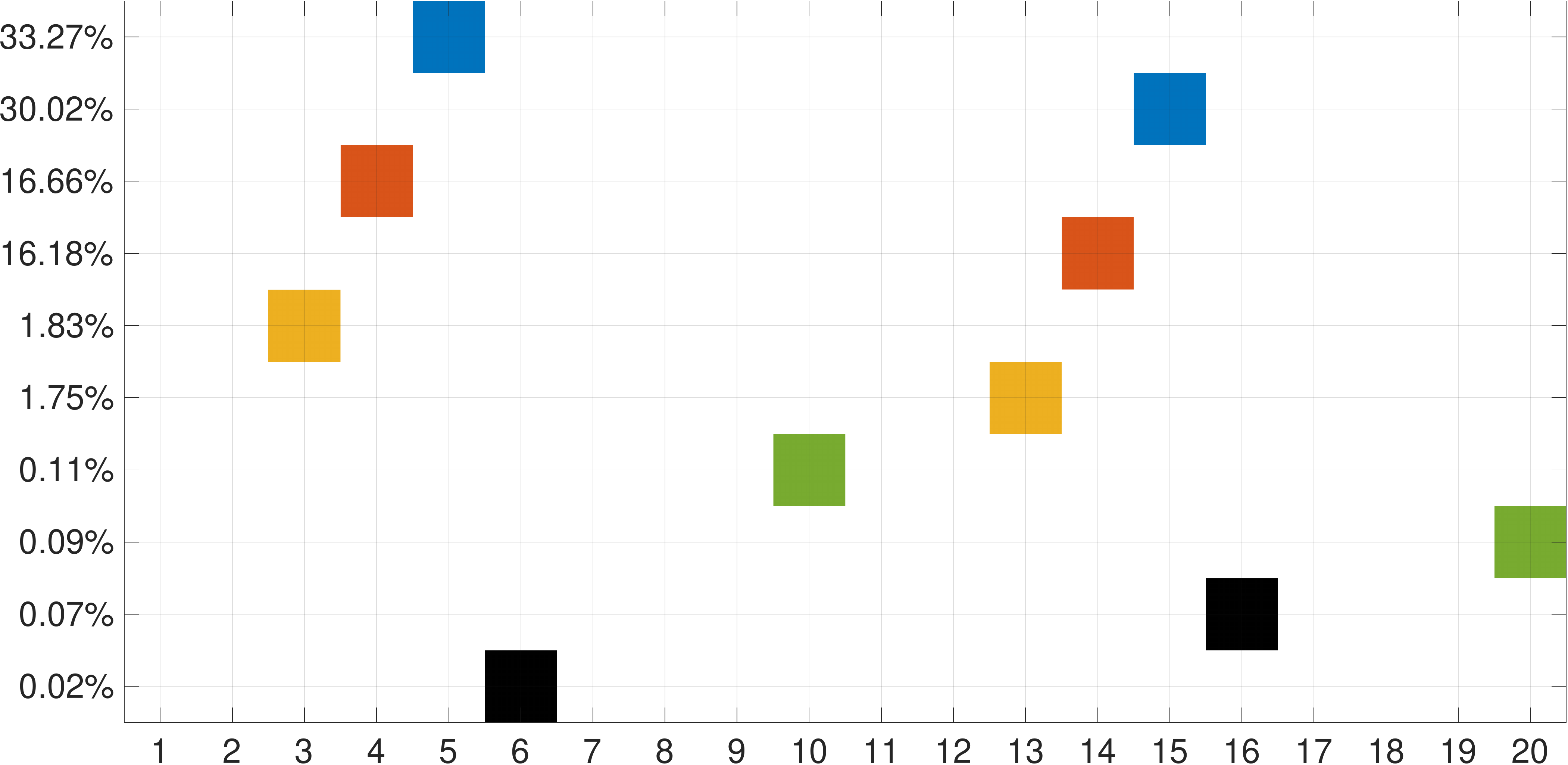}

	\caption{Each row of the matrix indicates the spatial support of a mode found by Algorithm~\ref{alg:sampling_optim} in Example 2 (the true locations are $5$ and $15$). The rows are ordered by the relative frequency (in $\%$), by which they are found and colored in such a way that it is easy to identify modes which are similar when locations $1$-$10$ and $11$-$20$ are switched.}
	\label{fig:1DExa2ModeFreq}
\end{figure}

We start with a toy problem where $m = 10$, $q = 20$, $d=t=1$ and the true unknown $\bfX$ is all zero except for $\bfX_5 = \bfX_{15} = 1$.
\paragraph{Example 1}
First, $\bfG$ is a random matrix constructed in the following way: its rows are drawn from a Gaussian distribution with zero mean and a block-diagonal covariance matrix $\mathbf{C} = {\rm blkdiag}\left(\mathbf{C}_1, \mathbf{C}_2\right)$, where $(\mathbf{C}_1)_{i,j} = 0.5^{|i-j|}$, $(\mathbf{C}_2)_{i,j} = 0.95^{|i-j|}$, $i,j = 1,\ldots, 10$. Then, each column is normalized to have unit $\ell_2$ norm.
Figure \ref{fig:1DExa1Setup} illustrates the set-up. Notice that due to the asymmetry in the design, the correct recovery of the source at location $15$ is more difficult due to the stronger correlation between columns $11$-$20$ of $\bfG$. We generate $\bfM$ by adding AWGN scaled by $20\% \norm{\bfM}_\infty$ to $\bfG \bfX$. We first run the MM Algorithm~\ref{alg:adpative_lasso} using a uniform initialization, \ie $\bfw^{(0)}_{i} = 1, \forall i=1\in [n]$, with $\lambda = 0.2\lambda_{max}$ where $\lambda_{max}= \max_{1 \leq i \leq n} \fronormsq{(\bfG^\top \bfM)_{[i]}}$ is the smallest regularization value for which no source is found as active using an $\ell_{2,1}$ regularization~\cite{Ndiaye_Fercoq_Gramfort_Salmon15,strohmeier-etal:16}. It recovers an $\bfX$ supported at locations $5$ and $11$, \ie it is not able to locate the second source correctly. Then, we run Algorithm~\ref{alg:sampling_optim} with $K = K_0 = 10\,000$, $K_{SC} = K_{SS} = 10$ and the same settings for the MM algorithm as before to obtain chains of posterior samples $\{\bfX^{(k)}\}_{k=1}^K$, and the corresponding posterior modes $\{\hat{\bfX}^{(k)}\}_{k=1}^K$. We clustered the modes based on their spatial support which reveals that a total of $16$ different modes were found. Figure \ref{fig:1DExa1ModeFreq} depicts the spatial support of the modes listed based on the relative frequency with which they were found. It reveals that, indeed, there is a larger uncertainty in the location of the second source (at true location $15$) and that in this scenario, the support of the mode which is found most often coincides with that of the true solution. To check that the MCMC sampler described in Algorithm~\ref{alg:sampling} is not simply stuck in this mode for a long time, we compute how many steps (with respect to index $k$) it takes on average before solutions $\{\hat{\bfX}^{(k)}\}_{k=1}^K$ change. The result is $1.63$ steps, which means that the sampler switches between modes very frequently and should be able to explore the posterior sufficiently well. In traditional UQ, the covariance matrix of the posterior samples $\{\bfX^{(k)}\}_{k=1}^K$ would be used to characterize uncertainty and correlation between activity at different locations. Here, we want to compare it to a matrix whose $(i,j)^{th}$ entry shows the relative frequency with which two locations $i$ and $j$ are simultaneously active in the support of the modes $\{\hat{\bfX}^{(k)}\}_{k=1}^K$. Such a matrix is another way to visualize the information given by Figure \ref{fig:1DExa1ModeFreq}. Figure \ref{fig:1DExa1CovVsModeMaps} shows that the covariance matrix reveals very little information about the true, sparse source locations and the larger ambiguity about the source at location $15$ induced by the asymmetric design of $\bfG$.
\paragraph{Example 2} While the posterior mode whose support coincided with that of the true solution was also found with the highest relative frequency, it is not clear whether this frequency is a reliable indication of the mode's true relative posterior mass. In general, this question is difficult to examine for high dimensional problems. Nonetheless, we constructed a second example to at least show that the frequencies are consistent: we now draw the rows of a $10 \times 10$ matrix $\tilde{\bfG}$ from a Gaussian distribution with zero mean and the covariance matrix $\mathbf{C}_2$ as for the previous example.
Then, we set $\bfG = [\tilde{\bfG}, \tilde{\bfG}]$, \ie the first and last $10$ columns of $\bfG$ are exactly the same. This means that the regression problem \eref{eq:FwdEq} and the posterior distribution are invariant with respect to switching the first and last $10$ entries. Every mode has a corresponding copy ``on the other side'', which should be found with the same relative frequency. All other settings are the same as in the previous example, except that we choose $\lambda = 0.5\lambda_{max}$ larger than before to boost modes which are only supported at a single location. Figure~\ref{fig:1DExa2ModeFreq} reveals that, indeed, all modes found are supported only at a single location and are found with a similar frequency as their corresponding copy. The average number of steps for the sampler to switch between different modes is now $1.40$ steps which is, again, very low. In addition, even the average number of steps it needs to switch between modes supported in locations $1$-$10$ to modes supported in $11$-$20$ is only $2.34$ steps.

\subsection{Simulated MEG data}
\begin{figure}[htp]
	\centering
	\includegraphics[clip,width=.9\columnwidth]{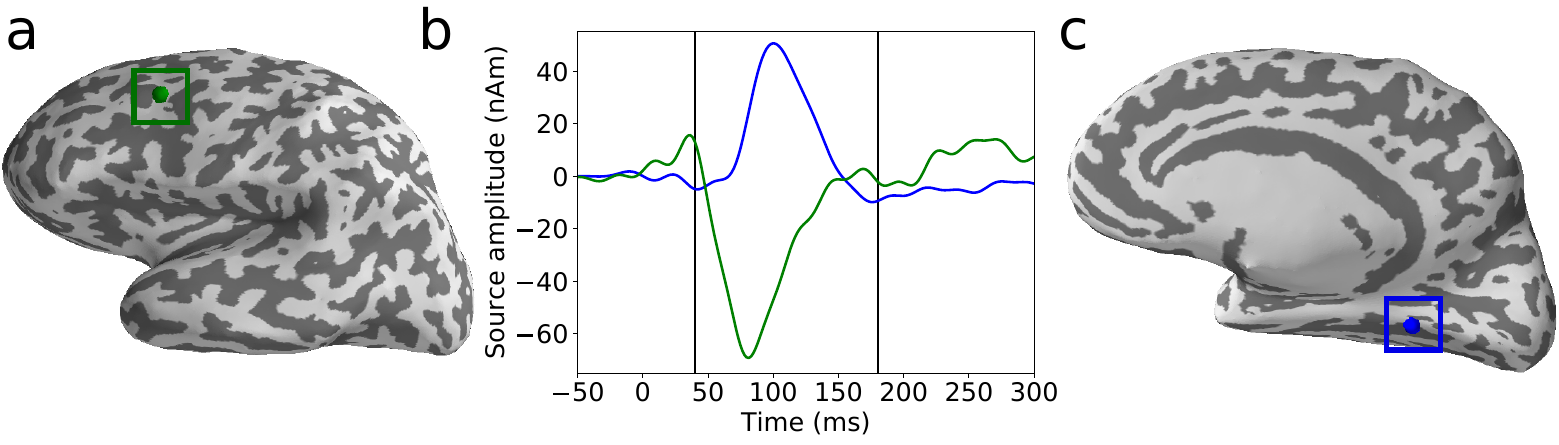}%

	\caption{Simulated MEG dataset. a) and c) show superficial and deep source (hidden in the medial view) locations, respectively. b) gives their corresponding waveforms color-coded by location.}
	\label{fig:simulated_data}
\end{figure}

We generated a realistic simulation based on a free-orientation ($d=3$) source model with $n=7498$ cortical locations and $m=306$ MEG sensors. Two of these locations were selected to be active, one in each hemisphere. One of the sources had a deep ventral location in the inferior occipital gyrus (Fig.~\ref{fig:simulated_data}-c), and the second one had a more superficial location in the motor cortex (Fig.~\ref{fig:simulated_data}-a). Their corresponding waveforms are shown in Fig.~\ref{fig:simulated_data}-b. When passed to the solvers, they are cropped between 40 to 180 ms to keep only the two peaks. This leads to $t=43$ time samples.

\begin{figure}[htp]
	\centering
	\includegraphics[width=\columnwidth]{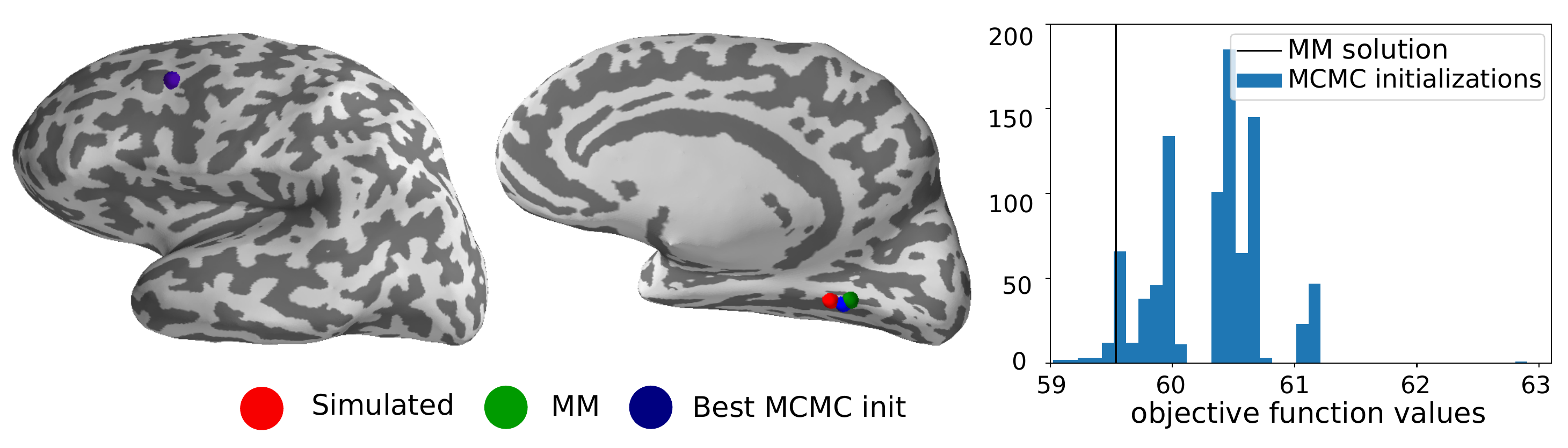}
	\caption{Location of simulated and estimated sources using the uniformly initialized MM solver (denoted as ``MM'') and best MCMC-based initialization in terms of objective function value. Left: estimation of the artificial source on the left hemisphere. Middle: estimation of the deep source on the right hemisphere. Right: histogram of the objective function value for 900 MCMC initializations (Algorithm~\ref{alg:sampling_optim}). The uniform initialization used for the MM (black vertical line) is not very bad, meaning that the basic MM is able to recover a good source estimates for some configurations. See Fig.~\ref{fig:hist_real_datasets} for a case where the basic MM fails.}
	\label{fig:simu_MM_best_MCMC}
\end{figure}

First, we re-visit the question whether we are able to find better source estimates using MCMC-derived initializations than with the uniformly initialized MM Algorithm~\ref{alg:adpative_lasso}, also in this high-dimensional example. For this, we first run the MM Algorithm~\ref{alg:adpative_lasso} using a uniform initialization, \ie $\bfw^{(0)}_{i} = 1, \forall i=1\in [n]$, with $\lambda = 0.05\lambda_{max}$.
Then, we run Algorithm~\ref{alg:sampling_optim} with $K_0 = 300$, $K = 900$, $K_{SC} = K_{SS} = 1$ and the same settings for the MM algorithm as before to obtain a chain of MM-optimized solutions $\{\hat{\bfX}^{(k)}\}_{k=1}^K$.
Figure~\ref{fig:simu_MM_best_MCMC}-c shows the histogram of the objective function values reached by these solutions (computed with Eq.~\eref{eq:lasso_new_gram}). The vertical black bar shows the value of the objective function of the uniformly initialized MM solver and we can see that some initializations indeed lead to source estimates with a lower objective value. Fig.~\ref{fig:simu_MM_best_MCMC}-a and Fig.~\ref{fig:simu_MM_best_MCMC}-b show the locations of the estimated sources resulting from uniform and best MCMC-based initialization. For the artificial source in Fig.~\ref{fig:simu_MM_best_MCMC}-a, both results find the exact location, so they are superposed. For the deeper source in Fig.~\ref{fig:simu_MM_best_MCMC}-b, neither result finds the exact position, but the MCMC-based initialization is closer. This means that the result did not only improve from an optimization point of view, but also judged by the quality criteria of the given application.

\begin{figure}[htp]
	\centering
	\includegraphics[clip,width=0.98\columnwidth]
{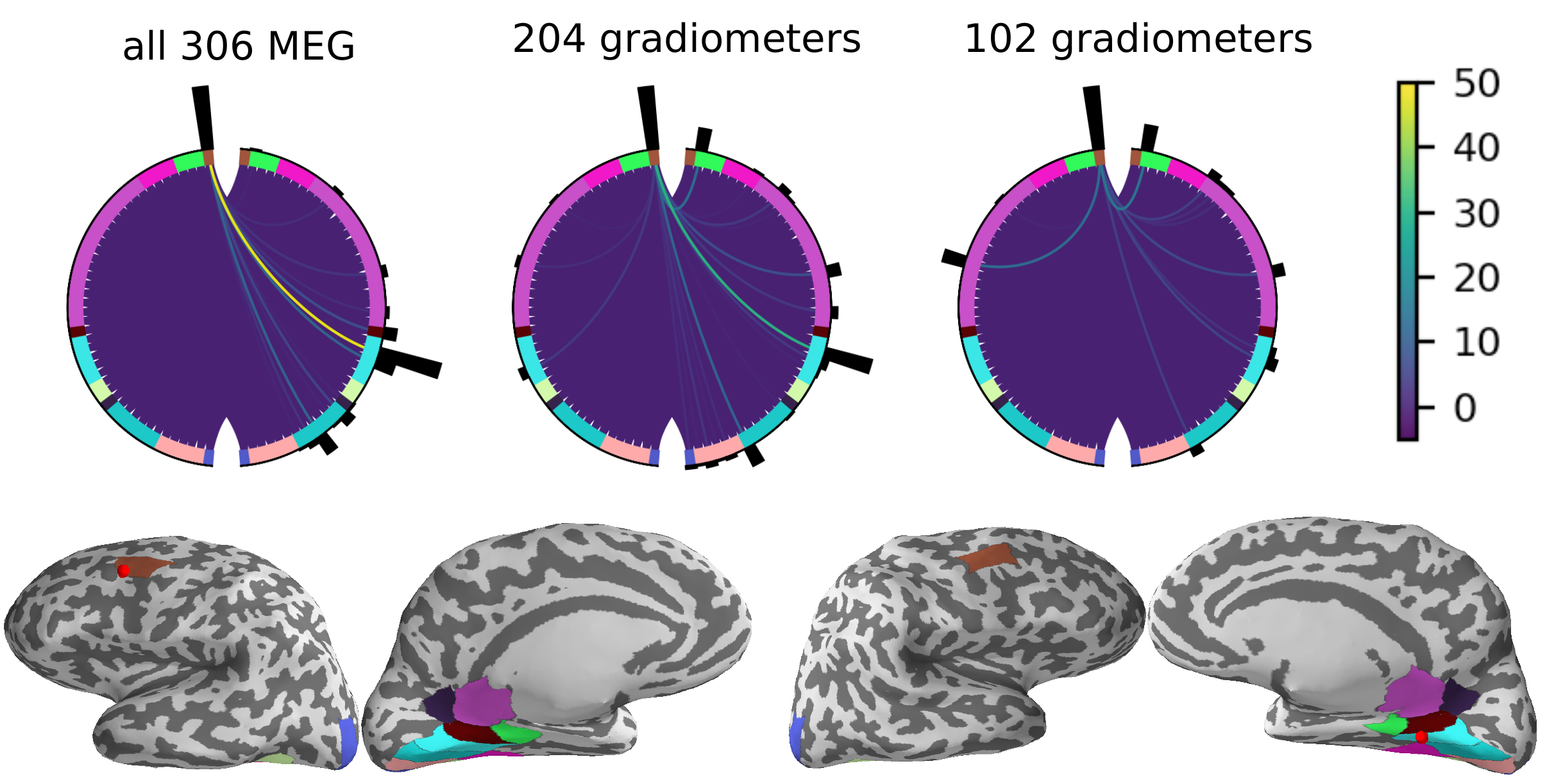}

	\caption{Source network analysis for simulated data: for a clearer presentation, the chain $\{\hat{\bfX}^{(k)}\}_{k=1}^{900}$ of modes was thinned to the 100 ones with the lowest objective function (Eq.~\eref{eq:L2pReg}). The first row of sub-figures displays the support of these best modes in the following way: each position in the circle represents a source location that was part of the support of at least one mode for one sensor configuration. The black bar attached to each position corresponds to the relative frequency with which this source location appeared as part of the support. Two positions are connected by a line if they were simultaneously part of the support and the color of this line corresponds to the relative frequency with which this happened. Note that the background of the circle is white, but it is densely covered by purple lines indicating rare connections. The positions are placed left or right, depending on which hemisphere they belong to. For symmetry, for each active source location, its counterpart on the other hemisphere was included in the graphic as well. In addition, the positions are grouped and colored based on a parcellation of the brain into anatomical regions (taken from an atlas). The second row of sub figures shows these regions in the brain and the simulated sources.}
	\label{fig:results_simu_circular}
\end{figure}

\begin{figure}[htp]
	\centering
	\begin{minipage}{\linewidth}
		\begin{minipage}{0.9\linewidth}
			\setlength\tabcolsep{0.1pt} 
			\includegraphics[width=\columnwidth]{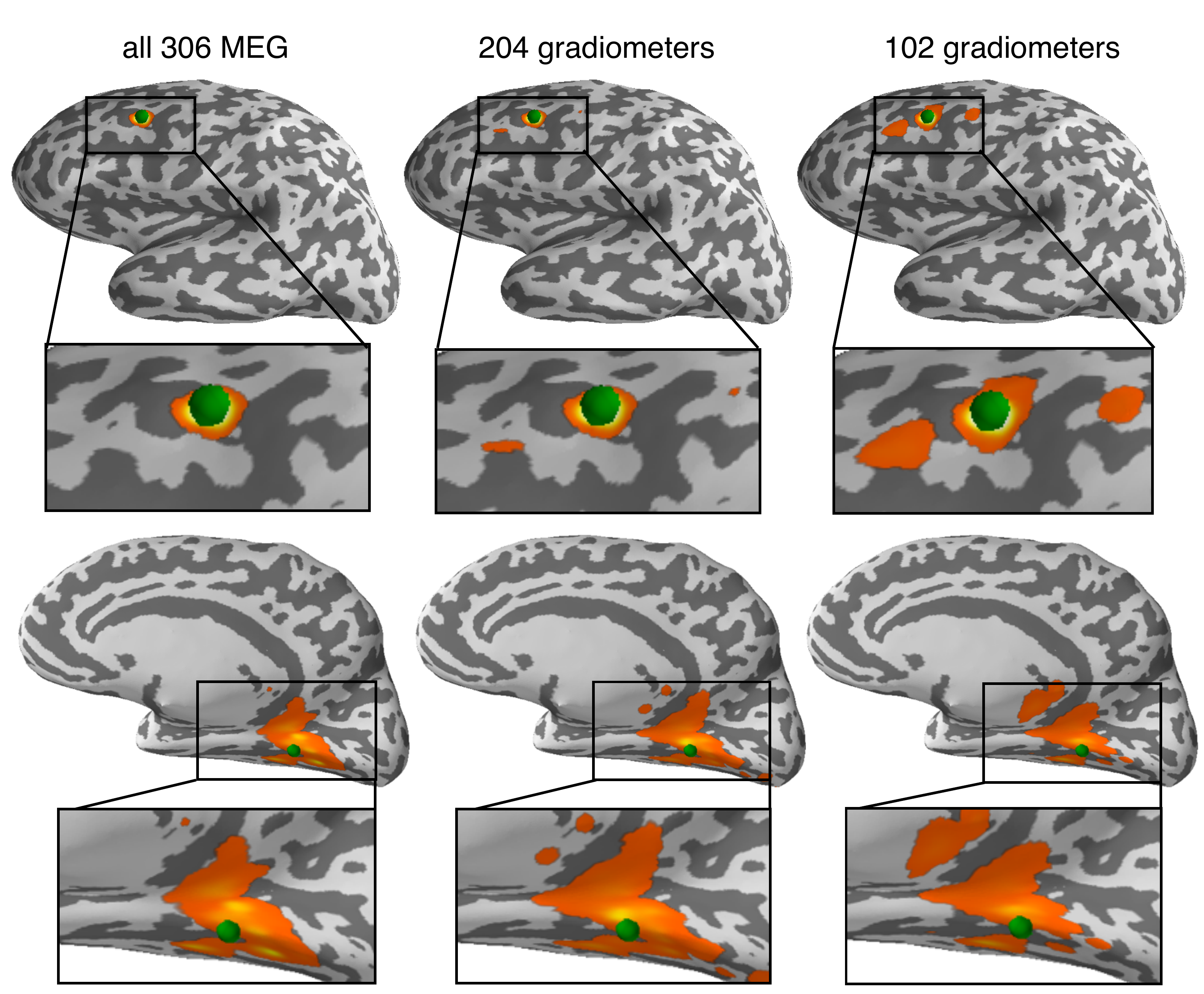}%
		\end{minipage}
		\begin{minipage}{0.09\linewidth}
				\includegraphics[width=\columnwidth]{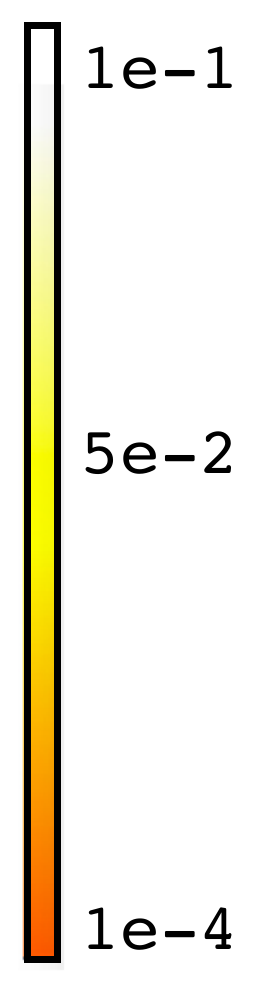}%
		\end{minipage}
	\end{minipage}
	\caption{The support of $\{\hat{\bfX}^{(k)}\}_{k=1}^{900}$ was extracted to build an uncertainty map by smoothing. The relative frequencies with which each source location was part of the support was computed and plotted on the brain surface together with the two simulated sources (green dots). Each column corresponds to the results for each of the three sensor setups examined. Less the number of sensors and/or more the source is deep, more uncertainty in the brain map. Note that the deep source is not in the support of the $\{\hat{\bfX}^{(k)}\}_{k=1}^{900}$, it seems to be recovered only due to smoothing.}
	\label{fig:results_simu_heat_maps}
\end{figure}

Now, we examine how the posterior modes found by Algorithm~\ref{alg:sampling_optim} react to changes in the measurement design. To do so, we switch from using all 306 MEG sensors to using only 204 gradiometers or each other gradiometer (102 sensors). By reducing the number of sensors we increase the under-determinedness of the problem, and the intuition is that it should lead to
more variability among the plausible sparse solutions. The graphical analysis, which is more involved for this high-dimensional scenario, is presented and described in Fig.~\ref{fig:results_simu_circular} and Fig.~\ref{fig:results_simu_heat_maps}. A first observation is that the superficial source in the premotor cortex was correctly identified as part of the support of every local minima when using the full 306 MEG sensors. It was however sometimes mis-localized when reducing the number of sensors (Fig.~\ref{fig:results_simu_circular}). A second observation is that the spatial spread of these miss-localizations is smaller for this superficial source than it is for the deep source. This deep source in the ventral cortex is more difficult to find even with all sensors. Indeed, none of the 100 best initialization perfectly localized the deep simulated source. In general, we can clearly see how the ambiguity increases when decreasing the number of sensors, and how the distribution of source networks gets more fuzzy. However, our analysis also provides useful local measures of these phenomena.

\subsection{Experimental MEG data}
We now repeat our analysis with two experimental open datasets. The first one is a recording of auditory evoked fields (MNE sample dataset~\cite{mne-python}). The second one contains visual evoked fields (visual condition of MNE sample dataset) for which source localization is a more difficult task due to the proximity between neural sources. The true nature of the underlying source network is also less clear for this second dataset.

\begin{figure}[htp]
	\centering
	\includegraphics[width=\columnwidth]{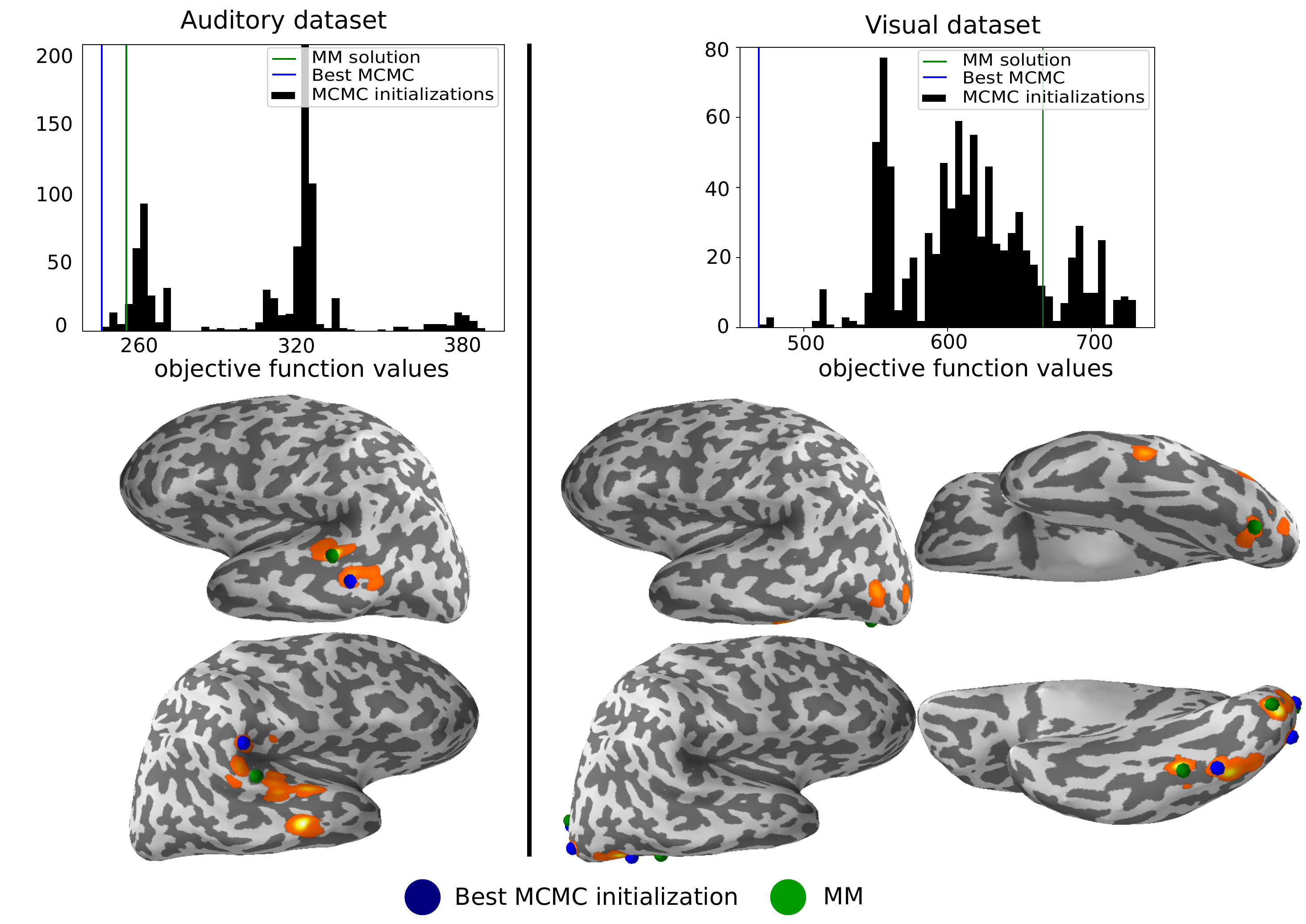}%
	\caption{Histogram of the objective function value of $\{\hat{\bfX}^{(k)}\}_{k=1}^{900}$ for auditory and visual datasets (306 MEG sensors). The histogram for visual dataset shows more MCMC initializations that outperform the uniform one in the MM solution. Under each histogram, these source configurations are shown on the left and right hemisphere.
	}
	\label{fig:hist_real_datasets}
\end{figure}

Figure~\ref{fig:hist_real_datasets} shows the equivalent to Fig.~\ref{fig:simu_MM_best_MCMC} for both datasets. Again, we see that lower objective function value can be obtained using MCMC-based initializations. The auditory sample dataset is commonly assumed to be generated by two bilateral focal sources around the auditory cortices in the superior temporal gyrus of the temporal lobe. Due to the superficial nature of these sources and their large distance, estimation of their position is regarded as a relatively simple task. Indeed, the histogram shows that using MCMC-based initializations does not help a lot to reduce the objective function compared to a uniformly initialized MM solution. However, in the case of the visual dataset, where several closed-by sources are active, the difference is quite drastic. The majority of the MCMC-based initializations lead to lower values of the objective function. Looking at the source distribution plots on the brain for both datasets, one can also observe more complex source configurations for the visual data.

\begin{figure}[htp]
	\centering
	\includegraphics[width=\columnwidth]{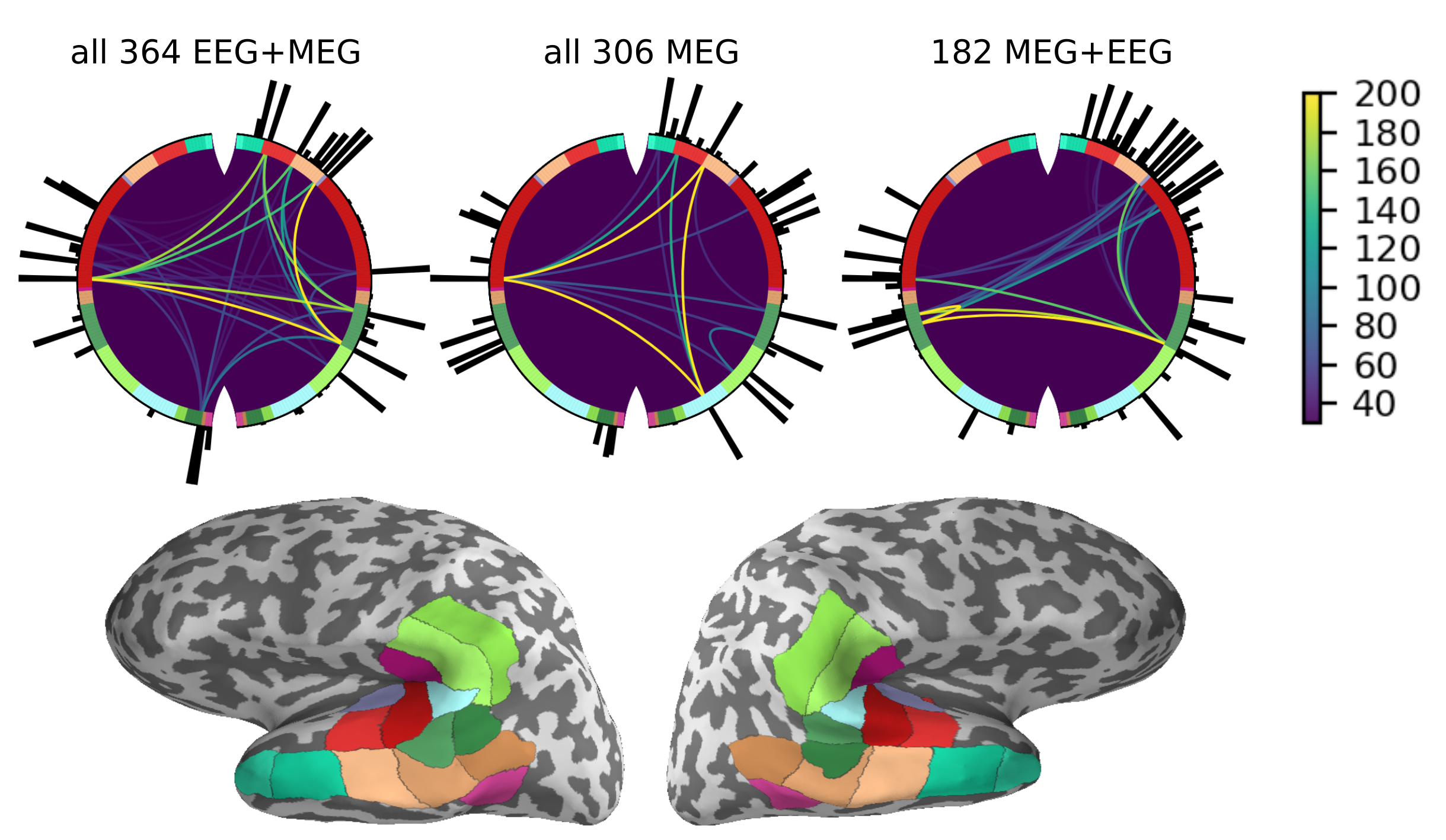}%
	\caption{Source network analysis for auditory data. The figures are constructed in the same way as described in Fig.~\ref{fig:results_simu_circular} except that all 900 mode samples are displayed.}
	\label{fig:circular_plots_LAud}
\end{figure}

\begin{figure}[htp]
	\centering
	\includegraphics[width=\columnwidth]{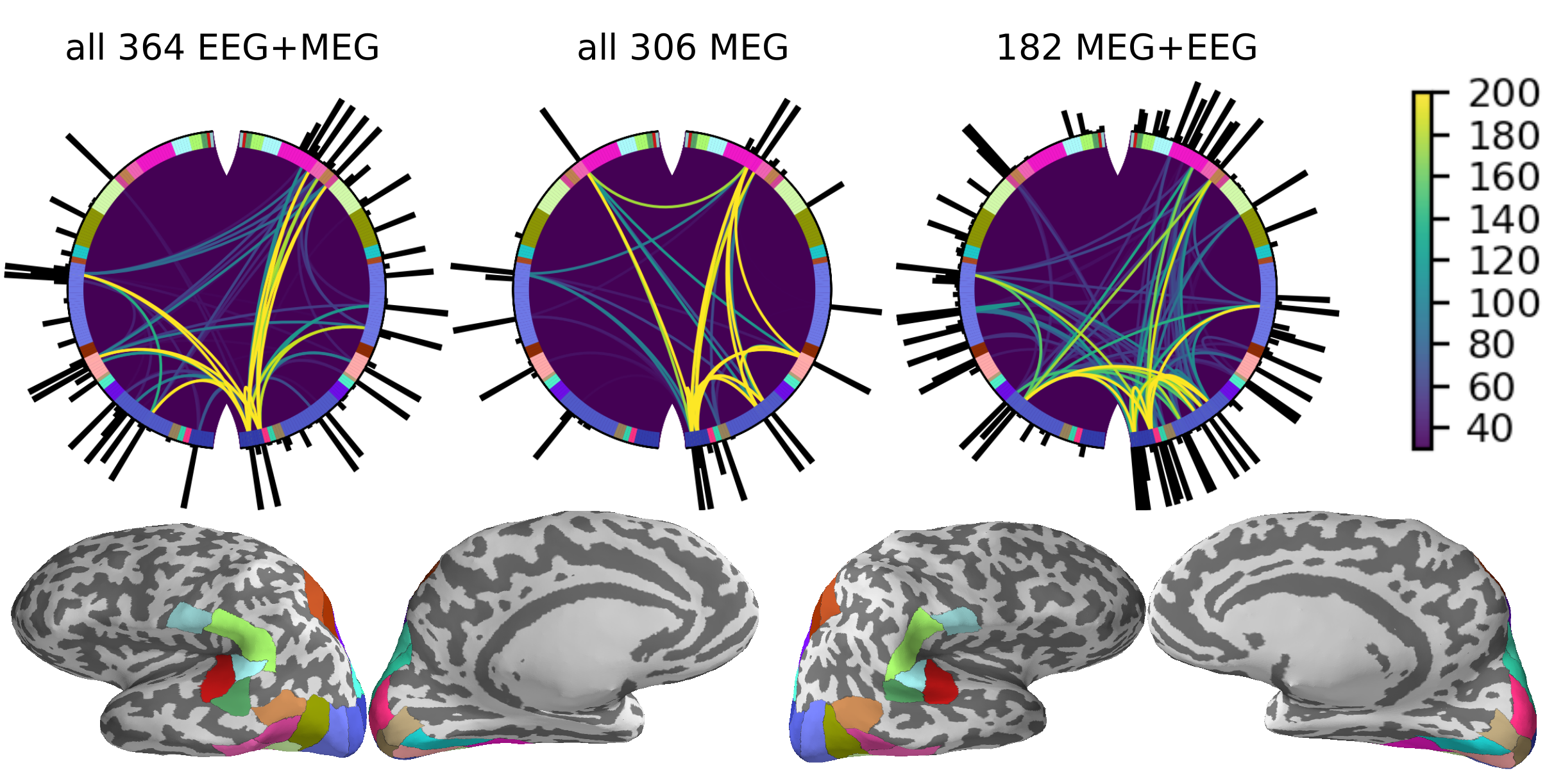}%
	\caption{Source network analysis for visual data. The figures are constructed in the same way as described in Fig.~\ref{fig:results_simu_circular} except that all 900 mode samples are displayed.}
	\label{fig:circular_plots_LVis}
\end{figure}

Next, we repeat the graphical source network analysis from Fig.~\ref{fig:results_simu_circular} for the two datasets. Figure~\ref{fig:circular_plots_LAud} shows the results for the auditory dataset and three sensor configurations: all 364 EEG + MEG sensors, all 306 MEG sensors or each other sensor resulting in 182 EEG + MEG sensors. One can see how adding EEG to MEG sensors reduces the ambiguity of the regression problem. The plots show less but more prominent modes, \ie the posterior mass is concentrated on fewer stable source configurations. We also see that the locations of the most prominent modes shift.
This is consistent with results of other studies on EEG-MEG combination \cite{MoStBrHa08,Lu14,AyVoKuHeKuGaHaWeKeRaWoHe14} as EEG is sensitive to some sources that MEG is almost blind to, \eg sources with a strong radial component. If we subsample the EEG+MEG sensors by only using every other location, the ambiguity and spatial spread of the recovered support increases. One can see that there is more activity in the dark green label, which corresponds to a brain area commonly not associated with auditory responses.
The connections between source locations show that none of the modes found really stands out, \ie is found much more often compared to the others. Most of the connections do not occur more than 200 times within the 900 samples, so they are part of the purple background of low frequency connections in the plots.\\
Figure~\ref{fig:circular_plots_LVis} shows the same results for the more complex visual dataset. Compared to the auditory dataset, we see that even with all sensors, the ambiguity of the regression problem seems to be a lot higher compared to the auditory dataset: we see that the posterior mass is distributed among many more source configurations. For the other two sensor configurations, we see similar effects as in the auditory data set. Nevertheless, it can be noticed that the large majority of identified sources with all MCMC initializations are on the right hemisphere. This is consistent with the known functional organization of the visual cortex. Indeed, in this experimental condition the subject was presented with checker board flashes on the left visual hemifield which is known to primarily project onto the right hemisphere of the cortex.

\section{Discussion}
\label{sec:Dis}

Scientific literatures relying either on frequentist or on Bayesian statistical inference often coexist in many fields ranging from machine learning, inverse problems, signal processing to computational biology.
In this work, we started from an under-determined, ill-conditioned MMV / multi-task regression problem and examined two seemingly unrelated approaches - MM as an optimization technique for tackling non-convex optimization problems arising in frequentist regression, and HBM as a Bayesian prior modeling framework. We showed that one obtains the same algorithms, and therefore the same solutions, when considering some specific choices of models, parameters and inference strategies. In particular the parallel was established between the $\ell_{2,1/2}$-norm regularized regression by MM and the full-MAP estimation for $\ell_{2,1}$ hierarchical priors with specific Gamma hyper-priors. We further showed that this conceptual parallel can be exploited to improve the MM solution by providing well-informed algorithmic initializations.



For this, we first constructed a multi-layered Gibbs sampler for the joint posterior density of our HBM.
Each sample is then used to initialize the MM step done with a state-of-the-art convex solver using block coordinate
descent techniques and acceleration strategies based on active sets. The sampler used has also an efficient sub-sampler for $\ell_{2,1}$ priors at its core. Despite the multi-modality of the posterior, the MCMC scheme is able to jump rapidly between the different attractors of the MM scheme. Indeed, using each sample as an initialization to the MM computation, one ends up in many different local minima /  modes of the posterior (cf. Fig~\ref{fig:hist_real_datasets}, Fig~\ref{fig:circular_plots_LVis}).
Therefore, this procedure allows us to reveal and explore different plausible source configurations in more details.
%

Based on this observation, we showcased how one can use the chain of local minima found by MCMC-initialized MM to analyze the variability of the different sparse solutions and how this yields different information compared to traditional and generic Bayesian uncertainty quantification techniques that use for example covariance estimates or credible sets derived from posterior samples~\cite{szabo2015frequentist}. It is also different from methods developed specifically for parametric M/EEG source localization based on dipole fitting~\cite{Fuchs20041442,Darvas2005355}. These latter approaches cannot easily be transferred to sparse, non-parametric approaches. On the other hand, our approach can easily be extended to include space-time-frequency structured sparsity constraints that can model more sophisticated source configurations than examined here \cite{Castano-Candamil-etal:2015}.
Using our developed techniques on simulations and actual data, one could observe that uncertainty in M/EEG is location specific and also source configuration specific. This is of course well-known by experts in this field, but here we provide a computational approach to visualize it and quantify it. This is an important incentive to develop such automated, data-dependent methods to quantify uncertainties in the context of M/EEG source imaging. In more conventional imaging methods such as computer tomography (CT) or magnetic resonance imaging (MRI), the signal originates from weak tissue interaction with strong external fields and the forward operator $\bfG$ depends almost exclusively on the physical properties of the scanner. In this situation, uncertainty is usually distributed in a smooth, well-known way over the image domain. Artifacts as well as real anatomical features are also easy to distinguish for a trained radiologist. The situation for M/EEG is very different. The weak signals originate from endogenous activity, and they are very dependent on dataset specific factors such as source orientation, location and attenuation which all depend on the geometry of the head of the analyzed subject. That is also why the forward matrix $\bfG$ needs to be constructed for each individual patient, after fixing the electrical properties of the head issues, which if wrong, increases the uncertainties.

When considering real data, the source to recover is often poorly understood, especially when it comes to pathological brain activity such as ictal or inter-ictal epileptic activity. In such a situation, providing a single source configuration as a result, together with an ad-hoc uncertainty quantification based on previous studies or acquired expertise, might not be an optimal use of the M/EEG data.
Instead, providing multiple hypotheses together, along with a quantification of their uncertainty, can be more useful. Indeed for applications such as pre-surgical epilepsy diagnosis, where M/EEG recordings are one of several diagnostic modalities, each candidate source configuration can provide some evidence for or against a diagnostic hypothesis that could lead to a surgery decision.
We therefore believe that future extensions of this work towards a consistent framework for interpreting and quantifying the multitude of potential results of sparse M/EEG source reconstruction approaches can have a significant impact on clinical settings.

\ack

The authors would like to thank Mathurin Massias for its valuable suggestions and readproofing of the manuscript. This work was supported by the French National Research Agency (ANR-14-NEUC-0002-01), the European Research Council Starting Grant SLAB ERC-YStG-676943 and in parts by the Engineering and Physical Sciences Research Council, UK (EP/K009745/1), the European Union's Horizon 2020 research and innovation programme H2020 ICT 2016-2017 under grant agreement No 732411 (as an initiative of the Photonics Public Private Partnership) and the Netherlands Organisation for Scientific Research (NWO 613.009.106/2383).


\appendix

\section{Slice-Within-Gibbs Sampler for Parameter Update}\label{sec:DetailsSliceSampler}

Within the Algorithm \ref{alg:sampling}, to update a group $\bfX_{[l]}$, we need to sample from the all the one-dimensional, SC densities
\begin{equation}
p_{post}(\bfX_{(i,j)}|\bfX_{-(i,j)}^{(k-1)},\bfM,\gamma^{(k)}), \qquad (i,j) \in [l] \enspace ,
\end{equation}
where $\bfX_{-(i,j)}$ refers all the coefficient of the matrix $\bfX$ except the term $(i,j)$.

To implement this efficiently, we can precompute several terms and make use of the specific spatio-temporal group structure of the posterior. We first derive the part of the likelihood \eref{eq:like} that depends on a given index pair $(i,j) \in [l]$:
\begin{eqnarray}
\fl \frac{1}{2} \fronormsq{\bfM - \bfG \bfX} = \sum_{j'}^{t} \frac{1}{2} \norm{\bfM_{(:,j')} - \bfG \bfX_{(:,j')}}_2^2 \proptoab{j} \\
\fl  \frac{1}{2} \norm{\bfM_{(:,j)} - \bfG \bfX_{(:,j)}}_2^2 = \frac{1}{2} \norm{\bfM_{(:,j)} - \left( \bfG_{(:, -i)} \bfX_{(-i,j)} + \bfG_{(:, i)} \bfX_{(i,j)} \right)}_2^2 \\
\fl \proptoab{i} \frac{1}{2} \norm{\bfG_{(:,i)}}_2^2 \, \bfX_{(i,j)}^2 + \bfG_{(:,i)}^T \left(\bfM_{(:,j)} - \bfG_{(:,-i)} \bfX_{(-i,j)} \right) \, \bfX_{(i,j)} \\
\fl = \frac{1}{2} \norm{\bfG_{(:,i)}}_2^2 \, \bfX_{(i,j)}^2 + \left( \left( \bfG^T \bfM\right)_{(i,j)} - \left({\bfG_{(:,i)}}^T \bfG\right) \bfX_{(:,j)} - \norm{\bfG_{(:,i)}}_2^2 \right) \, \bfX_{(i,j)}\\
\fl := a z^2 + b z \enspace , \quad \text{with} \quad z := \bfX_{(i,j)} \enspace , \quad a := \frac{1}{2} \norm{\bfG_{(:,i)}}_2^2 \enspace , \\
\fl b :=  \left( \bfG^T \bfM\right)_{(i,j)} - \left({\bfG_{(:,i)}}^T \bfG\right) \bfX_{(:,j)} - \norm{\bfG_{(:,i)}}_2^2
\end{eqnarray}
Note that $\norm{\bfG_{(:,i)}}_2^2 $ and $ \left( \bfG^T \bfM\right)$ can be precomputed. The challenging part in the computation of $b$ is to compute $\left({\bfG_{(:,i)}}^T \bfG\right)$, as one typically does not want to pre-compute the $q \times q$ matrix $G^T G$ and hold it in memory. However, to update all the $td$ components of the $l$-th group (\eg, in the visual evoked fields example, $t = 211$, $d = 3$) one only needs the $d \times q$ matrix $\left({\bfG_{(:,[l])}}^T \bfG\right)$. Thus, we compute $\left({\bfG_{(:,[l])}}^T \bfG\right)$ at the start of updating group $\bfX_{[l]}$ and hold it memory throughout the bloc update. Besides this, the most costly operation to compute $b$ is a dot product of vectors of size $q$. Next, we derive the part of the prior \eref{eq:condprior} that depends on $\bfX_{(i,j)}, (i,j) \in [l]$:
\begin{eqnarray}
\fl \sum_{l=1}^n \left( \frac{\fronorm{\bfX_{[l]}}^p}{\gamma_l} + \frac{d t}{p} \log(\gamma_l)\right) \quad \proptoab{\bfX_{(i,j)}, (i,j) \in [l]} \quad \gamma_l^{-1} \fronorm{\bfX_{[l]}}^p =  \gamma_l^{-1} \left( \sum_{(i',j') \in [l]} \bfX_{(i',j')}^2  \right)^{p/2}  \nonumber \\
\fl = \gamma_l^{-1} \left(\bfX_{(i,j)}^2 + \sum_{(i',j') \in [l] \atop (i',j') \neq (i,j)} \bfX_{(i',j')}^2  \right)^{p/2} =: c \left(z^2 + d \right)^{p/2} \enspace ,
\end{eqnarray}
with $c$ and $d$ defined and computed in an obvious way. Taken together, to update $\bfX_{(i,j)}$, we have to sample from the one-dimensional density:
\begin{equation}
p(z) \propto \exp \left(- a z^2 - b z \right) \exp \left(- c \left(z^2 + d \right)^{p/2} \right) =: p_1(z) p_2(z) \enspace . \label{eq:SCdensity}
\end{equation}
We take advantage of the fact that \eref{eq:SCdensity} factorizes in a Gaussian likelihood part $p_1(z)$ and a symmetric, log-concave prior part $p_2(z)$, and use a generalized form of slice sampling \cite{Ne03,RoCa05} as described in more detail in \cite{Lu16} and summarized in Algorithm \ref{alg:Slice}. Determining $S_2^y$ in our case is trivial:
\begin{equation}
\fl p_2(z) \geqslant y \quad \Leftrightarrow \quad c \left(z^2 + d \right)^{p/2} \leqslant - \log(y) \quad \Leftrightarrow |x| \leqslant \left( \left( \frac{- \log(y)}{c} \right)^{2/p} - d \right)^{1/2}
\end{equation}
Then, we use a slightly modified, more robust, version of the fast table-based algorithm described in \cite{Ch12} to sample from the truncated Gaussian distribution $p_1(z) \Indicator_{S^y_2}(z)$. As initialization for $z$, we always chose the current value of $\bfX_{(i,j)}$.

{\fontsize{4}{4}\selectfont
\begin{algorithm}[t]
\SetKwInOut{Input}{input}
\SetKwInOut{Init}{init}
\SetKwInOut{Parameter}{param}
\caption{\textsc{Slice Sampling}}
\Input{$p(z) \propto p_1(z) p_2(z), z, K_{(SS)} $}
\For{
     $k = 1$ to $K_{SS}$
    }
    {
    Draw $y$ uniform from $\left[0,p_2(z)\right]$ (vertical move).\\
    Determine $S_2^y \mydef \left\lbrace z \;|\; p_2(z) \geqslant y \right\rbrace$\\
	Draw $z$ from $p_1(z) \Indicator_{S^y_2}(z)$ (weighted horizontal move).
    }
{\Return{$z$} as a sample of $p(z)$}
\label{alg:Slice}
\end{algorithm}
}

%
%
%
%

\section{Accept-Reject Sampler for Hyperparameter Update} \label{sec:DetailsGammaSampler}

\begin{figure}[tb]
   \centering
\includegraphics[width = 0.8\textwidth]{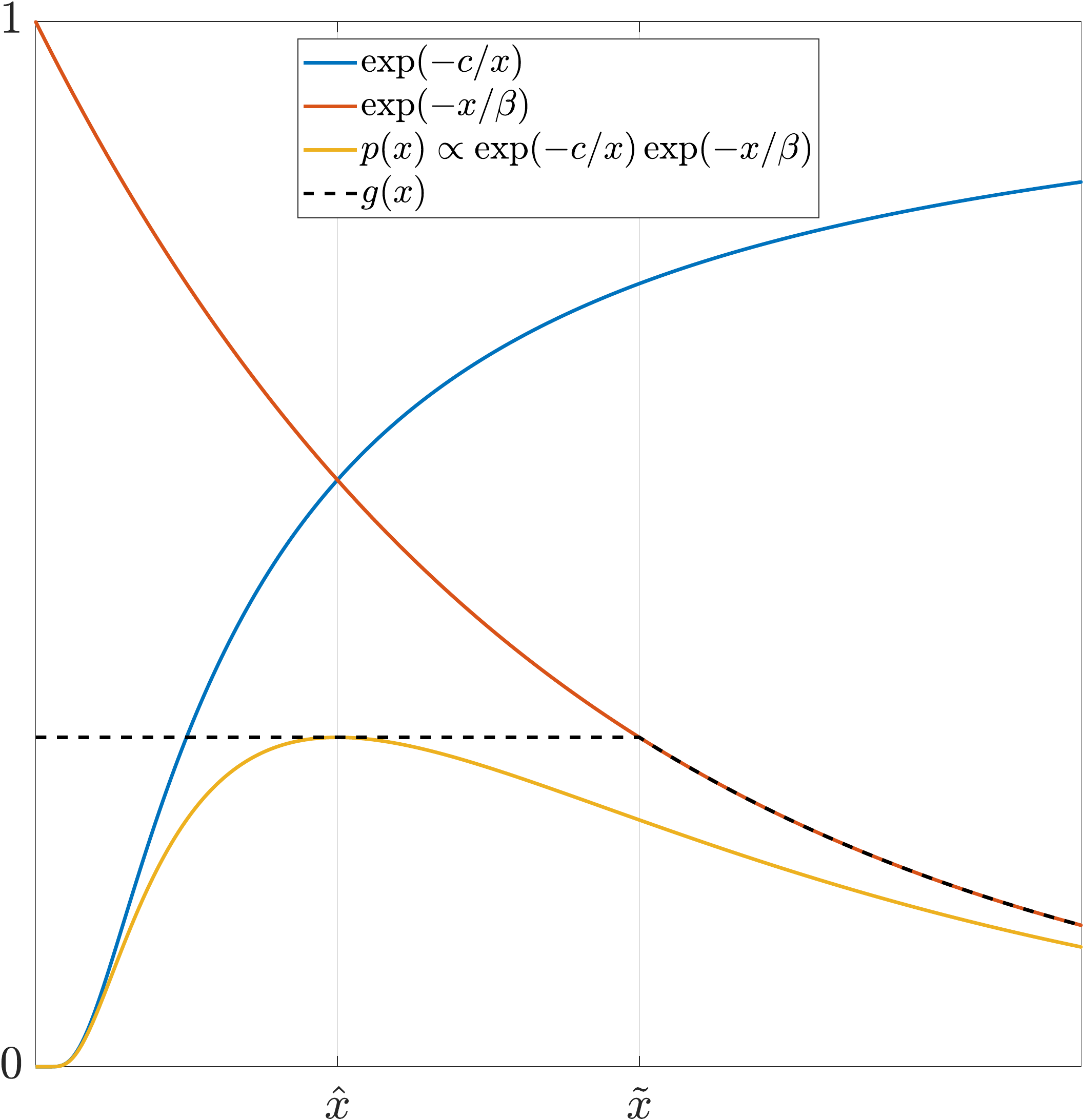}
\caption{Sketch of the quantities used in the accept-reject sampling Algorithm~\ref{alg:acceptreject}.}
   \label{fig:AccRejSketch}
\end{figure}

The conditional density \eref{eq:conPostGammaiRed} is of the type
\begin{equation}
\fl \qquad p(x) \propto \exp \left(- \frac{c}{x} \right) \exp \left(- \frac{x}{\beta} \right) \enspace, \qquad c, \beta \geqslant 0.
\end{equation}
Note that the first factor is monotonically increasing with limit $0$ for $x \searrow 0$ and limit $1$ for $x \rightarrow \infty$ while the second factor is proportional to a simple exponential distribution (\lcf Figure \ref{fig:AccRejSketch}). We can therefore easily construct a dominating density $g(x) \geqslant p(x)$ to carry out accept-reject sampling \cite[Section 2.3.2]{RoCa05} to generate a sample $z \sim p$: we generate $y \sim g$, $u \sim \mathcal{U}_{[0,1]}$ and accept $z = y$ if $u \leqslant p(y)/g(y)$ and repeat otherwise. Choosing $g(x) = \exp \left(- x/\beta\right)$ would yield a valid sampling density but this choice becomes inefficient with increasing $c$. Therefore, we split the sampling density into two parts:
\begin{equation}
g(x) = \cases{
\hat p & \text{if} $x < \tilde{x}$ \\
\exp \left(- \frac{x}{\beta} \right)	& \text{otherwise}}  \enspace,\label{eq:ARsamplingDensity}
\end{equation}
where $\hat p = \max_x p(x)$ is the maximal probability attained at $\hat x = \argmax_x p(x) = \sqrt{\beta c}$ and $\tilde{x} = \beta c/\hat x + \hat x$ is the solution to $\exp \left(- x/\beta\right) = \hat p$ (\lcf Figure \ref{fig:AccRejSketch}). Sampling from \eref{eq:ARsamplingDensity} is then straight-forward using $v, w \sim \mathcal{U}_{[0,1]}$: if one computes
\begin{equation}
\fl G_{x\geqslant \tilde{x}} = \int_{\tilde{x}}^\infty g(x) \, dx = \beta \exp(- \tilde{x}/\beta), \qquad
G_{x < \tilde{x}} = \int_0^{\tilde{x}} g(x) \, dx = \hat{p} \tilde{x} \enspace,
\end{equation}
then $v <  G_{x\geqslant \tilde{x}} / (G_{x\geqslant \tilde{x}} + G_{x < \tilde{x}})$ determines that we are in the tail, $x > \tilde{x}$, where we can use a simple inverse cumulative distribution method to draw a proposal from $g(x)$ using $w$. If $v$ determines that we are in $x \leqslant \tilde{x}$, then $x = w \tilde{x}$ is the proposal. For numerical precision, we only compute logarithms of probabilities and use that for $a > 0$, b$\geqslant 0$:
 \begin{equation}
 \log \left(a +b \right) = \log a + \log \left(1 + \exp \left(b - a \right)\right) \enspace .
 \end{equation}
 The whole sampling scheme is shown in Algorithm~\ref{alg:acceptreject}. We found the scheme to be efficient enough for all of our computations, \ie the chosen $g(x)$ is close enough to $p(x)$ to result in an accepted sample after a few trails. If this would become a problem, it would be easy to adaptively improve the dominating density.

{\fontsize{4}{4}\selectfont
\begin{algorithm}[t]
\SetKwInOut{Input}{input}
\SetKwInOut{Init}{init}
\SetKwInOut{Parameter}{param}
\caption{\textsc{Accept-Reject Algorithm for Hyperparameter Update}}
\Input{$c \geqslant 0, \beta > 0$}
Set $\hat x = \sqrt{\beta c}$.\\
Set $\log \hat p = -c/\hat x - \hat x/\beta$.\\
Set $\tilde{x} = \beta c/\hat x + \hat x$.\\
Set $\log G_{x\geqslant \tilde{x}} = \log \beta - \tilde{x}/\beta$.\\
Set $\log G_{x < \tilde{x}} = \log \hat p + \log \tilde{x}$. \\
Set $\log G_{tot} = \log G_{x\geqslant \tilde{x}} + \log\left(1 + \exp\left(G_{x < \tilde{x}} - G_{x\geqslant \tilde{x}}  \right)\right)$.\\
\While{true}
    {
    Draw $u, v, w$ uniform from $\left(0,1\right)$.\\
    \If{$\log v + \log G_{tot} < \log G_{x\geqslant \tilde{x}}$}{
		Set $W = \log w - \tilde{x}/\beta$.\\
		Propose $x = - \beta W$:\\
		\If{$\beta \log(u) < c/W$}{\Return{$x$} (acceptance)}
    }
    \Else{Propose $x = w \tilde{x}$:\\
    		\If{$\log u + \log \hat{p} < -c/x - x/\beta$}{\Return{$x$} (acceptance)}
    		}
    }
\label{alg:acceptreject}
\end{algorithm}
}

\clearpage

\section*{References}

\bibliographystyle{jphysicsB}
\bibliography{../LpHypermodelsEMEG_ref}

\end{document}